\documentclass[12pt]{article}
\pdfoutput=1
\usepackage{amsmath,amssymb,amsfonts}
\usepackage{a4wide,epsfig,psfrag,scalefnt}
\usepackage[dvipsnames]{xcolor}
\usepackage[pdftitle={Heavy-to-light form factors to three loops},
  pdfauthor={Matteo Fael, Tobias Huber, Fabian Lange, Jakob M\"uller, Kay Sch\"onwald, and Matthias Steinhauser},
  pdfkeywords={},
  bookmarks=true, linktocpage,
  colorlinks=true, allbordercolors=white, allcolors=blue]{hyperref}

\usepackage{braket}
\usepackage{placeins}
\usepackage{breqn}
\usepackage{slashed}
\usepackage{enumitem}
\usepackage[numbers,sort&compress]{natbib}
\usepackage{caption}
\usepackage{subcaption}
\usepackage{verbatim}

\graphicspath{ {figs/},{graphs/},{plots/veF1/},{plots/teF2/} }
\newcommand{\ep}{\epsilon}

\def\bsgamma{$\bar B \to X_{s} \, \gamma$}
\renewcommand{\b}{\beta}
\newcommand{\lb}{L_\beta}

\allowdisplaybreaks

\begin{document}

\title{\vskip-3cm{\baselineskip14pt
    \begin{flushright}
     \normalsize
     CERN-TH-2024-064,
     P3H-24-036,
     PSI-PR-24-13, \\
     SI-HEP-2024-14,
     TTP24-017,
     ZU-TH 28/24
    \end{flushright}} \vskip1.5cm
  Heavy-to-light form factors to three loops
}

\author{
  Matteo Fael$^{a}$,
  Tobias Huber$^{b}$,
  Fabian Lange$^{c,d}$,
  \\
  Jakob M\"uller$^{b}$,
  Kay Sch\"onwald$^{c}$,
  Matthias Steinhauser$^{e}$
  \\[1.0em]
  {\small\it (a)  Theoretical Physics Department, CERN,}\\
  {\small\it 1211 Geneva, Switzerland}
  \\
  {\small\it (b)
   Theoretische Physik 1, Center for Particle Physics Siegen (CPPS), Universit\"at Siegen,} \\
   {\small\it Walter-Flex-Stra{\ss}e 3, D-57068 Siegen, Germany
  }
  \\
  {\small\it (c) Physik-Institut, Universit\"at Z\"urich, Winterthurerstrasse 190,}\\
  {\small\it 8057 Z\"urich, Switzerland}
  \\
  {\small\it (d) Paul Scherrer Institut, 5232 Villigen PSI, Switzerland}
  \\
  {\small\it (e) Institut f{\"u}r Theoretische Teilchenphysik,
    Karlsruhe Institute of Technology (KIT),}\\
  {\small\it 76128 Karlsruhe, Germany}
}

\date{}

\maketitle

\thispagestyle{empty}

\begin{abstract}

  We compute three-loop corrections of
  $\mathcal{O}(\alpha_{s}^3)$ to form factors with one massive
  and one massless quark coupling to an external vector, axialvector,
  scalar, pseudoscalar, or tensor current.
  We obtain analytic results for the color-planar contributions, for the contributions of light-quark loops, and the contributions with two heavy-quark loops.
  For the computation of the remaining
  master integrals we use the ``expand and match'' approach which
  leads to semi-analytic results for the form factors. We implement
  our results in a {\tt Mathematica} and a {\tt Fortran} code which
  allows for fast and precise numerical evaluations in the physically
  relevant phase space. The form factors are used to compute the hard matching coefficients in Soft-Collinear Effective Theory for all currents. The tensor coefficients at light-like momentum transfer are used to extract the hard function in $\bar B \to X_s \gamma$ to three loops.

\end{abstract}


\newpage


\section{Introduction}

Form factors are the basic building blocks of scattering amplitudes in quantum field theories.  Most prominently, they represent the bulk of virtual corrections to physical observables.  The form factors for two massless external particles coupling to an external current have been computed up to four-loop order in QCD and QED for various combinations of particles and currents~\cite{Kramer:1986sg,Matsuura:1987wt,Matsuura:1988sm,Gehrmann:2005pd,Baikov:2009bg,Gehrmann:2010ue,Lee:2010ik,Gehrmann:2010tu,Henn:2016men,vonManteuffel:2016xki,Henn:2016wlm,Lee:2017mip,Lee:2019zop,vonManteuffel:2019wbj,vonManteuffel:2019gpr,vonManteuffel:2020vjv,Lee:2021uqq,Lee:2022nhh,Chakraborty:2022yan}. The heavy quark form factors, i.e.\ two fermions with the same mass coupling to a current, were known at the two-loop level for a long time~\cite{Barbieri:1972as,Barbieri:1972hn,Mastrolia:2003yz,Bonciani:2003ai,Bernreuther:2004ih,Bernreuther:2004th,Bernreuther:2005rw,Bernreuther:2005gw,Gluza:2009yy,Henn:2016tyf,Ahmed:2017gyt,Ablinger:2017hst,Lee:2018nxa} and partial three-loop results became available over the last decade~\cite{Henn:2016tyf,Lee:2018rgs,Ablinger:2018yae,Ablinger:2018zwz,Lee:2018nxa,Blumlein:2019oas,Blumlein:2023uuq}. Recently, the three-loop corrections for the vector, axialvector, scalar, and pseudoscalar currents were completed semi-analytically~\cite{Fael:2022rgm,Fael:2022miw,Fael:2023zqr}.

The heavy-to-light form factors of a heavy and a light fermion are especially
relevant for decays of heavy quarks such as $t \to b W^*$, $b \to c W^*$, and
$b \to u W^*$ or the production of a single top quark through the $t$-channel
process.  Specializing to QED, they also contribute to the muon decay in the
Fermi theory, see e.g.\ Refs.~\cite{Berman:1962uvx,Anastasiou:2005pn}. For some of the applications, neglecting the mass of the light fermion is a good first approximation in which the form factors were known to two-loop order for some time~\cite{Bonciani:2008wf,Asatrian:2008uk,Beneke:2008ei,Bell:2008ws}.\footnote{See also Refs.~\cite{Bell:2006tz,Bell:2007tv,Huber:2009se} for the computation  of the respective master integrals.}  Only a few years ago the full mass dependence of the heavy-to-light form factor became available at $\mathcal{O}(\alpha_{\text{s}}^2)$~\cite{Chen:2018dpt,Engel:2018fsb}.
Neglecting the light fermion mass, the color-planar corrections
at $\mathcal{O}(\alpha_{\text{s}}^3)$ to the vector, axialvector, scalar, and
pseudoscalar form factors were computed recently~\cite{Chen:2018fwb,Datta:2023otd}.

In this paper we compute the three-loop
corrections to the heavy-to-light form factors in full QCD, still
neglecting the mass of the light fermion.  We reproduce the analytic
results of Ref.~\cite{Datta:2023otd} in the color-planar limit and
extend it to the tensor form factors.  Furthermore, we provide
analytic results for the contributions of light-fermion loops and the contributions with two heavy-fermion loops for all
form factors.  For the remaining color factors we present
semi-analytic results in terms of expansions around kinematic points
following the strategy of Ref.~\cite{Fael:2021kyg} which was already
applied to the three-loop corrections to the massive form factors in
Refs.~\cite{Fael:2022rgm,Fael:2022miw,Fael:2023zqr}.  We restrict
ourselves to the physically interesting regions relevant for the
heavy-fermion decay and the heavy-fermion production in the
$t$-channel.  Furthermore, we present results for generic external
currents.  The specification to vertices appearing in the Standard
Model or other theories of interest is straightforward. We provide our
analytic results in the form of ancillary files accompanying this
paper, and the numeric results for the full form factors as
\texttt{Mathematica} and \texttt{Fortran} programs which perform an
interpolation based on a dense grid~\cite{FFh}.

The QCD form factors can be used to compute the hard matching coefficients to Soft-Collinear Effective theory (SCET)~\cite{Bauer:2000yr,Bauer:2001yt,Beneke:2002ph,Beneke:2002ni} at leading power in the SCET expansion. The infrared  divergences still present in the QCD form factors are removed during the procedure of infrared subtraction, yielding finite SCET matching coefficients. While their one-loop expressions have been computed in the founding SCET papers (see also Ref.~\cite{Bosch:2004th}), the two-loop coefficients for the vector and axial vector current were computed in Refs.~\cite{Bonciani:2008wf,Asatrian:2008uk,Beneke:2008ei,Bell:2008ws}. In Ref.~\cite{Bell:2010mg} the results were extended to the scalar and tensor currents. In the present paper the matching coefficients are computed to three-loop order for all currents considered.

An immediate application of the matching coefficients of the tensor current at light-like momentum transfer concerns the inclusive decay $\bar B \to X_s \gamma$. In a SCET-based approach, the decay rate is formulated in a factorized form as the product of a hard function times a convolution of a jet with a soft function~\cite{Korchemsky:1994jb,Akhoury:1995fp,Neubert:2004dd}. While the latter two are known to three loops already~\cite{Becher:2005pd,Becher:2006qw,Bruser:2018rad,Bruser:2019yjk}, the hard function has to date only been evaluated to two loops~\cite{Ali:2007sj,Ligeti:2008ac,Dehnadi:2022prz}. With the three-loop matching coefficients at hand we close this gap and compute the three-loop QCD correction to the hard function in $\bar B \to X_s \gamma$.

In the recent study~\cite{Dehnadi:2022prz}, the authors claim the performance
of a next-to-next-to-next-to-leading-logarithmic analysis of the photon energy spectrum in $\bar B \to X_s \gamma$ including three-loop corrections to the renormalization-scale independent part of the hard, jet, and soft functions in SCET (i.e.\ a study to N${}^3$LL${}^\prime$ accuracy). However, for the hard function this piece has only become available with the calculation presented here. In Ref.~\cite{Dehnadi:2022prz} the missing numerical coefficient at three loops was treated as a nuisance parameter. Our explicit three-loop calculation shows that the exact numerical value of the parameter in question lies more than a factor of two outside the variation region assumed in Ref.~\cite{Dehnadi:2022prz}.

The remainder of this paper is structured as follows: In
Section~\ref{sec::form-factors} we introduce the form factors and
discuss their renormalization, the infrared subtraction, as well as
the Ward identities of the currents which relate some of the form
factors.  Our calculational strategy is described in
Section~\ref{sec::technicalities}.  We then present our results and
discuss the analytic and numeric results in Section~\ref{sec::results}
and~\ref{sec::results_num}, respectively. The hard function in $\bar B \to X_s \gamma$ is presented in Section~\ref{sec:hardfunctionbsgamma}.
We conclude in Section~\ref{sec::conclusion}. In the Appendixes we present explicit results for  the projectors to all form factors. Furthermore, we describe the
program {\tt FFh2l} where our results are implemented and which
allows for a fast and precise numerical evaluation.


\section{Form factors}
\label{sec::form-factors}

\subsection{Currents and form factors}

The theoretical framework used for our calculation is QCD supplemented with
external currents formed by a heavy ($Q$) and a light quark field
($q$). In this paper we consider the vector, axialvector, scalar, pseudoscalar, and tensor currents
\begin{eqnarray}
  j_\mu^v &=& \bar{\psi}_Q\gamma_\mu\psi_q\,,\nonumber\\
  j_\mu^a &=& \bar{\psi}_Q\gamma_\mu\gamma_5\psi_q\,,\nonumber\\
  j^s &=& \bar{\psi}_Q\psi_q\,,\nonumber\\
  j^p &=& \mathrm{i} \bar{\psi}_Q\gamma_5\psi_q\,, \nonumber\\
  j_{\mu\nu}^t &=& \mathrm{i} \bar{\psi}_Q \sigma_{\mu\nu} \psi_q\,,
  \label{eq::currents}
\end{eqnarray}
where $\sigma_{\mu\nu} = {\rm i}[\gamma^\mu,\gamma^\nu]/2$ is
anti-symmetric in the indices $\mu$ and $\nu$.
The wave functions of the heavy and light quark fields are denoted by
$\psi_Q$ and $\psi_q$, respectively.
We use the currents from Eq.~(\ref{eq::currents}) to construct vertex
functions $\Gamma(q_1, q_2)$ via
\begin{equation}
  \int \frac{\mathrm{d}^4 y}{(2\pi)^4} \, \mathrm{e}^{\mathrm{i} q \cdot y} \langle \psi^{\text{out}}_Q(q_2, s_2)|j^x(y)|\psi^{\text{in}}_q(q_1, s_1)\rangle = \bar{u}(q_2, s_2) \Gamma(q_1, q_2) u(q_1, s_1) \delta^{(4)}(q-q_1-q_2)\,,
  \label{eq:current-to-vertex}
\end{equation}
which are independent of the spin indices $s_1$ and $s_2$ and which can be decomposed into scalar form factors.
We follow the notation introduced in Ref.~\cite{Fael:2022miw}
and define them as
\begin{eqnarray}
  \Gamma_\mu^v(q_1,q_2) &=&
  F_1^v(q^2)\gamma_\mu
  - \frac{\rm i}{m}F_2^v(q^2) \sigma_{\mu\nu} q^\nu
  + \frac{2}{m} F_3^v(q^2) q_{\mu}
  \,, \nonumber\\
  \Gamma_\mu^a(q_1,q_2) &=&
  F_1^a(q^2)\gamma_\mu\gamma_5
  - \frac{{\rm i}}{m} F_{2}^a(q^2) \sigma_{\mu\nu}q^\nu \gamma_5
  + \frac{2}{m}F_{3}^a(q^2) q_\mu \gamma_5
  \,, \nonumber\\
  \Gamma^s(q_1,q_2) &=& F^s(q^2)
  \,, \nonumber\\
  \Gamma^p(q_1,q_2) &=& {\rm i} F^p(q^2) {\gamma_5}
  \,, \nonumber\\
  \Gamma_{\mu\nu}^t(q_1,q_2) &=&
  \mathrm{i} F^t_1(q^2) \sigma_{\mu\nu}
  + \frac{F^t_2(q^2)}{m} \left( q_{1,\mu} \gamma_\nu - q_{1,\nu} \gamma_\mu \right)
  + \frac{F^t_3(q^2)}{m} \left( q_{2,\mu} \gamma_\nu - q_{2,\nu} \gamma_\mu \right)
  \nonumber\\&&\mbox{}
  + \frac{F^t_4(q^2)}{m^2} \left( q_{1,\mu}q_{2,\nu} - q_{1,\nu}q_{2,\mu} \right)
  \,.
  \label{eq::Gamma}
\end{eqnarray}
Here, $q_1$ is the incoming momentum of the massless quark and
$q_2$ is the outgoing momentum of the heavy quark. Furthermore, we
have $q=q_1-q_2$, with $q^2=s$, $q_1^2=0$, and $q_2^2=m^2$. In all vertex functions
the colour structure is a simple Kronecker delta
in the fundamental colour indices of the external
quarks and is not written out explicitly.

For the perturbative expansion of the scalar form factors
we introduce
\begin{eqnarray}
  F &=& \sum_{i\ge0} \left(\frac{\alpha_s(\mu)}{\pi}\right)^i F^{(i)}
        \,,
        \label{eq::Fas}
\end{eqnarray}
where $\alpha_s$ depends on the number of active flavours.
We will use $\alpha_s^{(n_l)}$ (with $n_f=n_l+n_h$)
for the parametrization of the
ultraviolet renormalized but still infrared divergent form factors
and for the finite matching coefficients
where also the  infrared divergences have been subtracted.
Here, $n_f$ is the number of active flavours, i.e., for the $b\to u$
vertex corrections we have $n_f=5$ with $n_h = 1$. The non-zero tree-level contributions are given by
\begin{eqnarray}
  F^{v,(0)}_1 = F^{a,(0)}_1 =F^{s,(0)} =F^{p,(0)} =F^{t,(0)}_1 =1\,.\qquad
\end{eqnarray}

The form factors of the heavy-light currents do not get
contributions from so-called singlet diagrams where the external
current couples to a closed quark loop.  This allows us to use
anti-commuting $\gamma_5$ without ambiguity. Since one of the quarks is massless it is
always possible to anti-commute $\gamma_5$ to one end of the fermion
string and obtain simple relations for the
axialvector and pseudoscalar form factors
to their vector and scalar counterparts.
In our case we have
\begin{align}
  F_1^a = F_1^v\,,\qquad F_2^a = F_2^v\,,\qquad
  F_3^a = F_3^v\,,\qquad F^s = F^p\,.
  \label{eq:chiral-symmetry}
\end{align}
We use these relations as internal cross-check for our calculation.

In the work \cite{Datta:2023otd} the vector and axialvector form
factors have been considered  with a slightly different decomposition of
the vertex functions. The authors have introduced scalar factors $G_1$, $G_2$
and $G_3$ which are related to ours via
\begin{align}
     F_1^v &= G_1 + \frac{1}{2} G_2 ~, &
     F_2^v &= -\frac{1}{2} G_2 ~, &
     F_3^v &= -\frac{1}{4} G_3 ~.
\end{align}

\subsection{Renormalization}

For the three-loop calculation of the form factors we have to perform
the standard parameter renormalization of the strong coupling and the quark masses, the wave function
renormalization of the massive and massless external quarks, and the
renormalization of the external currents.  Furthermore, we decouple
the contribution from the heavy quark from the running of $\alpha_s$.
Then the combination with the subtraction terms from the infrared
divergences is more convenient.
We thus write the ultraviolet renormalized form factors as
\begin{equation}
  F^{x} = Z_{x} \left(Z_{2,Q}^{\mathrm{OS}}\right)^{1/2} \left(Z_{2,q}^{\mathrm{OS}}\right)^{1/2} F^{x,\rm bare} \Bigg|_{\alpha_s^{\rm bare}=Z_{\alpha_s}\alpha_s^{(n_f)} ,\, m^{\rm bare}=Z_m^{\rm OS}m^{\rm OS}, \, \alpha_s^{(n_f)} = \zeta_{\alpha_s}^{-1} \alpha_s^{(n_l)}} \,.
  \label{eq:renormalization}
\end{equation}
The bare one-loop vertex corrections develop $1/\epsilon^2$ terms
and at two-loop order we even have quartic poles. Thus the
(on-shell) renormalization and decoupling constants
are required to order $\epsilon^4$ at one-loop order
and to order $\epsilon^2$ at two loops.

Let us summarize the renormalization constants appearing in Eq.~(\ref{eq:renormalization}), up to which orders they are needed, and which schemes we choose:
\begin{itemize}
\item The renormalization of $\alpha_s$ is needed to two-loop order and is performed in the $\overline{\rm MS}$ scheme~\cite{Jones:1974mm,Caswell:1974gg,Egorian:1978zx}.
\item The renormalization of the heavy-quark mass $m$ is required to two-loop order.
  We choose the on-shell scheme~\cite{Tarrach:1980up,Gray:1990yh}, in which we need the one-loop result to order $\epsilon^4$ and the two-loop result to order $\epsilon^2$~\cite{Chetyrkin:1999ys,Chetyrkin:1999qi,Melnikov:2000qh,Marquard:2016dcn}.
\item The on-shell wave function renormalization constant of the
  heavy quark, $Z_{2,Q}^{\mathrm{OS}}$, is needed to three-loop order and can
  be found in Refs.~\cite{Broadhurst:1991fy,Melnikov:2000zc,Marquard:2007uj}.
  Again, we need the one-loop result to order $\epsilon^4$ and the two-loop result to order $\epsilon^2$~\cite{Marquard:2018rwx}.
\item The wave function renormalization constant of the light quark, $Z_{2,q}^{\mathrm{OS}}$,
  starts at order $\alpha_s^2$ and is needed up to
  three-loop order~\cite{Chetyrkin:1997un}.
  We need the two-loop result to order $\epsilon^2$~\cite{Gerlach:2018hen}.
\item Since the vector and axialvector current are conserved, their anomalous dimensions vanish and we have $Z_v = Z_a = 1$.
\item The anomalous dimension of the scalar and pseudoscalar currents corresponds to the anomalous dimension of the quark mass and we thus have $Z_s = Z_p = Z_m$, which we need to three loops.
  We choose to renormalize it both in the $\overline{\rm MS}$ as well as in the on-shell scheme.
  $Z_m^{\overline{\rm MS}}$ is available from Refs.~\cite{Tarrach:1980up,Tarasov:1982plg,Larin:1993tq}.
  For $Z_m^{\rm OS}$, we again need the one-loop result to order $\epsilon^4$ and the two-loop result to order $\epsilon^2$~\cite{Tarrach:1980up,Gray:1990yh,Chetyrkin:1999ys,Chetyrkin:1999qi,Melnikov:2000qh,Marquard:2016dcn}.
\item The tensor current has a non-vanishing anomalous dimension which cannot be deduced from other quantities.
  We need it to three loops to construct $Z_t$ in the $\overline{\rm MS}$ scheme~\cite{Broadhurst:1994se,Gracey:2000am}.
\item Finally, we decouple the heavy quark(s) from the running by employing the decoupling relation $\alpha_s^{(n_f)} = \zeta_{\alpha_s}^{-1} \alpha_s^{(n_l)}$, where we remind the reader that $n_f=n_l+n_h$.
  We require the decoupling relation to two loops~\cite{Weinberg:1980wa,Ovrut:1980uv,Ovrut:1981ue,Wetzel:1981qg,Bernreuther:1981sg,Bernreuther:1983zp,Chetyrkin:1997un}, the one-loop result to order $\epsilon^4$, and the two-loop result to order $\epsilon^2$~\cite{Chetyrkin:1997un,Schroder:2005hy,Chetyrkin:2005ia,Grozin:2007fh,Grozin:2011nk,Gerlach:2018hen}.
\end{itemize}

\subsection{Ward identities}

Using the equations of motion, one can derive the Ward identities
\begin{equation}
  \begin{split}
    \partial^\mu j_\mu^v &= \mathrm{i} m \,j^s \,, \\
    \partial^\mu j_\mu^a &= m \,j^p
  \end{split}
\end{equation}
between the renormalized vector and scalar as well as between the axialvector and pseudoscalar currents.
The equations of motion imply that both the mass and the currents are renormalized in the on-shell scheme.
Due to Eq.~(\ref{eq:chiral-symmetry}) it is sufficient to consider the vector and the scalar currents in the following.
Employing Eq.~(\ref{eq:current-to-vertex}), we can rewrite the Ward identity as
\begin{eqnarray}
  {-} q^\mu \Gamma_\mu^v = m \, \Gamma^s
\end{eqnarray}
on the level of the renormalized vertices (see, e.g., Ref.~\cite{Bonciani:2008wf}).
Using Eq.~(\ref{eq::Gamma}) then leads to the relation
\begin{eqnarray}
  F_1^v {-} \frac{2s}{m^2} F_3^v = F^s
  \label{eq:Ward-identity}
\end{eqnarray}
between the renormalized form factors.
This provides an important check on our results later, which we discuss in Section~\ref{sec::results_num}.

\subsection{\label{sec::IR}Infrared subtraction and matching onto SCET}

Infrared singularities of multi-leg QCD amplitudes with a massive and massless partons has been discussed in Refs.~\cite{Becher:2009kw,Liu:2022elt}.
By specifying ourselves to the case $Q \to q$, i.e.\ one massive initial quark and one massless final state quark, we can write the $Z$ factor associated to the
infrared subtraction in the minimal scheme in the following way:
\begin{align}
 \ln Z &= \frac{\alpha_s}{4 \pi}
          \bigg[ \frac{\Gamma_0'}{4 \ep^2}  + \frac{\Gamma_0}{2 \ep} \bigg]
       + \left(\frac{\alpha_s}{4 \pi} \right)^2
       \bigg[ - \frac{3 \beta_0 \Gamma_0'}{16 \ep^3}
       + \frac{\Gamma_1' - 4 \beta_0 \Gamma_0}{16 \ep^2} + \frac{\Gamma_1}{4 \ep} \bigg]
       \notag \\
       &+ \left(\frac{\alpha_s}{4 \pi} \right)^3
       \bigg[ \frac{11 \beta_0^2 \Gamma_0'}{72 \ep^4} - \frac{5 \beta_0 \Gamma_1'
       + 8 \beta_1 \Gamma_0' - 12 \beta_0^2 \Gamma_0}{72 \ep^3}
        + \frac{\Gamma_2' - 6 \beta_0 \Gamma_1 - 6 \beta_1 \Gamma_0}{36 \ep^2}
        + \frac{\Gamma_2}{6 \ep} \bigg]
        \notag \\
       &+ \mathcal{O}(\alpha_s^4) ,
\label{eq::Z}
\end{align}
where $\alpha_s \equiv \alpha_s^{(n_l)}(\mu)$,
\begin{equation}
    \Gamma = \gamma^Q(\alpha_s)+\gamma^q(\alpha_s)
    -\gamma^{\mathrm{cusp}}(\alpha_s) \log \left( \frac{\mu}{m (1 - x)}\right)
    = \sum_{n=0}^{\infty}
    \Gamma_n
    \left(\frac{\alpha_s}{4 \pi} \right)^{n+1}
\end{equation}
with $x=s/m^2$ and
\begin{equation}
    \Gamma'=\frac{\partial}{\partial \log \mu} \Gamma = -\gamma^{\mathrm{cusp}}(\alpha_s).
\end{equation}
The coefficients in the perturbative series of the light-like cusp anomalous dimension
\begin{equation}
\gamma^{\mathrm{cusp}} (\alpha_s) =
\sum_{n=0}^\infty
\gamma^{\mathrm{cusp}}_n
\left(\frac{\alpha_s}{4 \pi} \right)^{n+1}
\end{equation}
are available up to four-loop order~\cite{Moch:2004pa,Henn:2016men,Davies:2016jie,Henn:2016wlm,Lee:2017mip,Moch:2017uml,Grozin:2018vdn,Moch:2018wjh,Lee:2019zop,Henn:2019rmi,vonManteuffel:2019wbj,Henn:2019swt,vonManteuffel:2020vjv,Agarwal:2021zft}. Up to three loops we have
\begin{align}
\gamma^{\mathrm{cusp}}_0 &= 4 C_F, \notag \\
\gamma^{\mathrm{cusp}}_1 &= 4 C_F
\left[
C_A \left( \frac{67}{9} - \frac{\pi^2}{3}\right)
-\frac{20}{9} T_F n_l
\right],
\notag \\
\gamma^{\mathrm{cusp}}_2 &= 4 C_F
\Bigg[
  C_A^2 \left( \frac{245}{6}-\frac{134 \pi ^2}{27}+\frac{22 \zeta_3}{3}+\frac{11 \pi ^4}{45} \right)
  +C_F T_F n_l \left( 16 \zeta_3-\frac{55}{3} \right)
  \notag \\ &
  + C_A T_F n_l \left( -\frac{418}{27}+\frac{40 \pi ^2}{27} -\frac{56 \zeta_3}{3}\right)
  -\frac{16}{27}T_F^2 n_l^2
\Bigg].
\end{align}
The perturbative expansion of the
anomalous dimension $\gamma^i$ (for $i=q,Q$)  can be written as
\begin{equation}
\gamma^i (\alpha_s)=
\sum_{n=0}^\infty
\gamma^i_n
\left(\frac{\alpha_s}{4 \pi} \right)^{n+1}
\end{equation}
and it can be extracted from the divergent part of the quark form factor.
$\gamma^q$ is know to four-loop order~\cite{Moch:2005id,Moch:2005tm,Baikov:2009bg,vonManteuffel:2020vjv,Agarwal:2021zft}; up to three loops the results read:
\begin{align}
    \gamma^q_0 &= -3 C_F ,\notag \\
    \gamma^q_1 &=
    C_A C_F \left(-\frac{961}{54}-\frac{11 \pi ^2}{6}+26 \zeta_{3}\right)
    +C_F^2 \left(-\frac{3}{2}+2 \pi ^2-24 \zeta_{3}\right)
    +C_F n_l T_F \left(\frac{130}{27}+\frac{2 \pi ^2}{3}\right), \notag \\
    \gamma^q_2 &=
    C_A^2 C_F
    \left(
        -\frac{139345}{2916}
        -\frac{7163 \pi ^2}{486}
        +\frac{3526 \zeta_{3}}{9}
        -\frac{83 \pi ^4}{90}
        -\frac{44 \pi ^2 \zeta_{3}}{9}
        -136 \zeta_{5}
    \right)
    \notag \\ &
    +C_A C_F^2
    \left(
        -\frac{151}{4}
        +\frac{205 \pi ^2}{9}
        -\frac{844 \zeta_{3}}{3}
        +\frac{247 \pi ^4}{135}
        -\frac{8 \pi ^2 \zeta_{3}}{3}
        -120 \zeta_{5}
    \right)
    \notag \\ &
    +C_F^3
    \left(
        -\frac{29}{2}
        -3 \pi ^2
        -68 \zeta_{3}
        -\frac{8 \pi ^4}{5}
        +240 \zeta_{5}
        +\frac{16 \pi ^2 \zeta_{3}}{3}
    \right)
    \notag \\ &
    +C_A C_F T_F n_l
    \left(
        -\frac{17318}{729}
        +\frac{2594 \pi ^2}{243}
        -\frac{1928 \zeta_{3}}{27}
        +\frac{22 \pi ^4}{45}
    \right)
    \notag \\ &
    +C_F^2 T_F n_l
    \left(
        \frac{2953}{27}
        -\frac{26 \pi ^2}{9}
        +\frac{512 \zeta_{3}}{9}
        -\frac{28 \pi ^4}{27}
    \right)
    +C_F  T_F^2 n_l^2
    \left(
        \frac{9668}{729}
        -\frac{40 \pi ^2}{27}
        -\frac{32 \zeta_{3}}{27}
    \right).
\end{align}
For massive quarks, $\gamma^Q$ is available up to three loops~\cite{Korchemsky:1987wg,Korchemsky:1991zp,Kidonakis:2009ev,Grozin:2014hna,Grozin:2015kna,Bruser:2019yjk}:
\begin{align}
    \gamma^Q_0 &= -2 C_F ,\notag \\
    \gamma^Q_1 &=
    C_A C_F \left(-\frac{98}{9}+\frac{2 \pi ^2}{3}-4 \zeta_{3}\right)
    +\frac{40 }{9}C_F T_F n_l ,
    \notag \\
    \gamma^Q_2 &=
    C_A^2 C_F
    \left(
        -\frac{343}{9}
        +\frac{304 \pi ^2}{27}
        -\frac{740 \zeta_{3}}{9}
        -\frac{22 \pi ^4}{45}
        -\frac{4 \pi ^2 \zeta_{3}}{3}
        +36 \zeta_{5}
    \right)
    \notag \\ &
    +C_A C_F T_F  n_l
    \left(
        +\frac{356}{27}
        -\frac{80 \pi ^2}{27}
        \frac{496 \zeta_{3}}{9}
    \right)
    +C_F^2 T_F  n_l
    \left(
        \frac{110}{3}
        -32 \zeta_{3}
    \right)
    +\frac{32}{27} C_F T_F^2 n_l^2 .
\end{align}

Note that $Z$ introduced in Eq.~(\ref{eq::Z}) is defined in terms of $\alpha_s^{(n_l)}$.
Thus, the decoupling relation has to be applied to the form factors in $d$ dimensions as discussed in the previous section.
We then have
\begin{eqnarray}
  C &=& Z^{-1} F\,, \label{eq:matching}
\end{eqnarray}
where $F$ is any of the ultraviolet renormalized form factors. The corresponding matching coefficient $C$ is finite (i.e., the limit $\epsilon\to0$ can be taken), and expanded perturbatively in analogy to Eq.~\eqref{eq::Fas}. Note that $C_3^t$ and $C_4^t$ vanish in four dimensions since the pseudotensor current is reducible in four space-time dimensions. This serves as another non-trivial check of our calculation.

Like $Z$, the matching coefficients $C$ are expanded in $ \alpha_s^{(n_l)}(\mu)$. They satisfy the renormalization group equation (RGE)
\begin{align}
    \frac{d}{d\ln(\mu)} \, C(s,\mu) = \left[\gamma^{\mathrm{cusp}} (\alpha_s^{(n_l)}) \, \ln\left(\frac{(1-x) m}{\mu} \right) + \gamma^H(\alpha_s^{(n_l)}) + \gamma^{\mathrm{QCD}}(\alpha_s^{(n_f)}) \right] \, C(s,\mu) \, , \label{eq:RGEC}
\end{align}
with $\gamma^H = \gamma^Q + \gamma^q$. The quantity $\gamma^{\mathrm{QCD}}$ is the anomalous dimension of the corresponding QCD current. It is expanded in $\alpha_s^{(n_f)}$ and can be extracted from the general formula~\cite{Gracey:2000am}\footnote{Note the typo in Eq.~(7) of Ref.~\cite{Gracey:2000am}: $-144 T_F^2 C_F^2$ should read $-144 T_F^2 n_f^2$, in accordance with Eq.~(6) of Ref.~\cite{Gracey:2000am}.}
\begin{align}
   \gamma_{(n)} =& - (n-1)(n-3) C_F \bigg(\frac{\alpha_s^{(n_f)}}{4\pi}\bigg) + \Big[ 4(n-15)T_F n_f + (18n^3 - 126n^2 + 163n + 291 )C_A
\nonumber \\
&  \hspace*{150pt}- 9(n-3)( 5n^2 - 20n + 1)C_F \Big] \,
\frac{(n-1)}{18} \, C_F \, \bigg(\frac{\alpha_s^{(n_f)}}{4\pi}\bigg)^2  \nonumber \\[1.0em]
& + \bigg\{ \Big[ 144n^5 - 1584n^4 + 6810n^3 - 15846n^2 + 15933n + 11413 \nonumber \\
&  \hspace*{30pt} - 216n(n-3)(n-4)(2n^2-8n+13)\zeta_3) \Big] C_A^2  \nonumber \\[0.6em]
& \hspace*{10pt}  - \Big[ 3 \, (72n^5 - 792n^4 + 3809n^3 - 11279n^2 + 15337n + 1161 )  \nonumber \\
&  \hspace*{30pt}  - 432n(n-3)(n-4)(3n^2-12n+19)\zeta_3 \Big]
C_A C_F  \nonumber \\[0.6em]
& \hspace*{10pt} - \Big[ 18(n-3)(17n^4 - 136n^3 + 281n^2 - 36n + 129)
 \nonumber \\
& \hspace*{30pt} + 864n(n-3)(n-4)(n^2 - 4n + 6)\zeta_3 \Big] C_F^2 \nonumber \\[0.6em]
& \hspace*{10pt} + \Big[ 8(3n^3 + 51n^2 - 226n - 278 ) + 1728(n-3)\zeta_3 \Big] \, C_A T_F n_f \nonumber \\[0.6em]
& \hspace*{10pt} - \Big[ 12(17n^3 + n^2 - 326n + 414) + 1728(n-3)\zeta_3 \Big] C_F T_F n_f \nonumber \\
& \hspace*{10pt} + 16(13n - 35) T_F^2 n_f^2 \bigg\}
\frac{(n-1)}{108} \, C_F \, \bigg(\frac{\alpha_s^{(n_f)}}{4\pi}\bigg)^3 + \mathcal{O}(\alpha_s^4) \,   \label{eq:gammaQCDgeneral}
\end{align}
via $\gamma^{\mathrm{QCD}}_{\{s,v,t\}} = -2 \gamma^{}_{\{(0),(1),(2)\}}$, where $\gamma_{(1)} = 0$ due to the conservation of the vector current.

The structure in Eq.~\eqref{eq:RGEC} allows us to distinguish two scales; the scale $\mu$ that governs the renormalization group evolution in SCET, and a second scale $\nu$ that governs the renormalization group evolution in QCD. The matching coefficients $C(s,\mu,\nu)$ then fulfil the two separate RGEs
\begin{align}
    \frac{d}{d\ln(\mu)} \, C(s,\mu,\nu) =& \left[\gamma^{\mathrm{cusp}} (\alpha_s^{(n_l)}(\mu)) \, \ln\left(\frac{(1-x) m}{\mu} \right) + \gamma^H(\alpha_s^{(n_l)}(\mu)) \right] \, C(s,\mu,\nu) \, , \label{eq:RGESCET}\\[0.5em]
    \frac{d}{d\ln(\nu)} \, C(s,\mu,\nu) =& \;\;\, \gamma^{\mathrm{QCD}}(\alpha_s^{(n_f)}(\nu))  \, C(s,\mu,\nu) \, . \label{eq:RGEQCD}
\end{align}
The dependence of the matching coefficients on $L_\mu = \ln(\mu^2/m^2)$ and $L_\nu = \ln(\nu^2/m^2)$ is then most conveniently derived by combining the running and the decoupling relation,
\begin{align}
\alpha_s^{(n_f)}(\nu) =& \alpha_s^{(n_f)}(\mu) \left[1- \beta_0^{(n_f)} \ln\Big(\frac{\nu^2}{\mu^2}\Big)
\Big(\frac{\alpha_s^{(n_f)}(\mu)}{4\pi}\Big) \right. \nonumber\\
& \left . +\Big({\beta_0^{(n_f)}}^2 \ln^2\Big(\frac{\nu^2}{\mu^2}\Big)-\beta_1^{(n_f)} \ln\Big(\frac{\nu^2}{\mu^2}\Big)\Big)\Big(\frac{\alpha_s^{(n_f)}(\mu)}{4\pi}\Big)^2 + \mathcal{O}(\alpha_s^3)\right] \, , \label{eq:run} \\[0.6em]
\alpha_s^{(n_f)}(\mu) =& \alpha_s^{(n_l)}(\mu) \left[1+ \frac{4}{3} L_\mu T_F \Big(\frac{\alpha_s^{(n_l)}(\mu)}{4\pi}\Big)\right. \nonumber\\
& \left . + \Big(\frac{16}{9} L_\mu^2 T_F^2 + C_F T_F (4 L_\mu + 15) +\frac{4}{9} C_A T_F (15 L_\mu -8)\Big)\Big(\frac{\alpha_s^{(n_l)}(\mu)}{4\pi}\Big)^2 + \mathcal{O}(\alpha_s^3) \right] \, . \label{eq:dec}
\end{align}
Note that contrary to Eq.~\eqref{eq:matching} the four-dimensional version of the decoupling relation is sufficient here. The coefficients of the QCD $\beta$ function follow from
\begin{align}
\frac{d \alpha_s^{(n_f)}(\mu)}{d\ln\mu} =& -2 \alpha_s^{(n_f)}(\mu) \left[ \beta_0^{(n_f)} \Big(\frac{\alpha_s^{(n_f)}(\mu)}{4\pi}\Big)
+\beta_1^{(n_f)} \Big(\frac{\alpha_s^{(n_f)}(\mu)}{4\pi}\Big)^2 + \mathcal{O}(\alpha_s^3)\right]
\end{align}
and assume their usual form
\begin{align}
  \beta_0^{(n_f)} =& \frac{11}{3} C_A - \frac{4}{3} n_f T_F \, , \\[0.6em]
  \beta_1^{(n_f)} =& \frac{34}{3} C_A^2 - \frac{20}{3} n_f T_F C_A - 4 n_f T_F C_F \, .
\end{align}


\section{Technicalities}
\label{sec::technicalities}

For our calculation we use the canonical chain based on
{\tt qgraf}~\cite{Nogueira:1991ex}, {\tt tapir}~\cite{Gerlach:2022qnc}, {\tt exp}~\cite{Harlander:1998cmq,Seidensticker:1999bb}, the in-house {\tt FORM}~\cite{Vermaseren:2000nd,Ruijl:2017dtg} code {\tt calc}, {\tt Kira}~\cite{Maierhofer:2017gsa,Klappert:2020nbg} and {\tt FireFly}~\cite{Klappert:2019emp,Klappert:2020aqs}.
All one- and two-loop and some of the three-loop master integrals
are computed to sufficiently high order in $\epsilon$ analytically. For the remaining three-loop master integrals we construct semi-analytic results based on
``expand and match''~\cite{Fael:2021kyg,Fael:2022rgm,Fael:2022miw,Fael:2023zqr}.

\subsection{Amplitude and projectors}

\begin{figure}[t]
	\begin{center}
		\setlength{\tabcolsep}{30pt}
		\begin{tabular}{ccc}
			\includegraphics[width=0.2\textwidth]{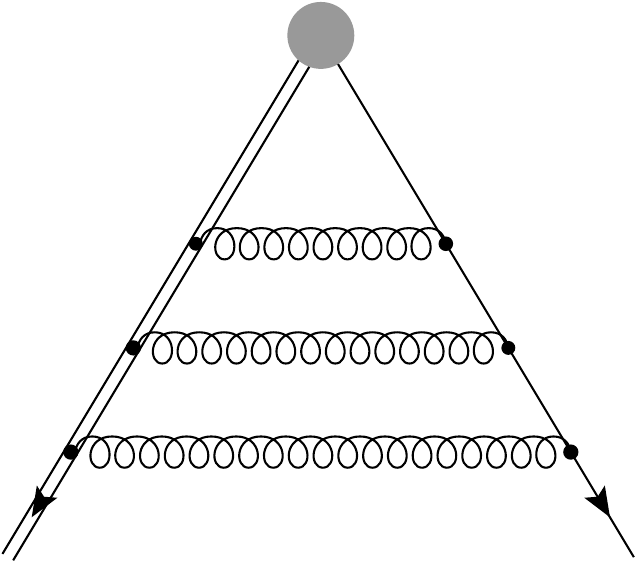}&
			\includegraphics[width=0.2\textwidth]{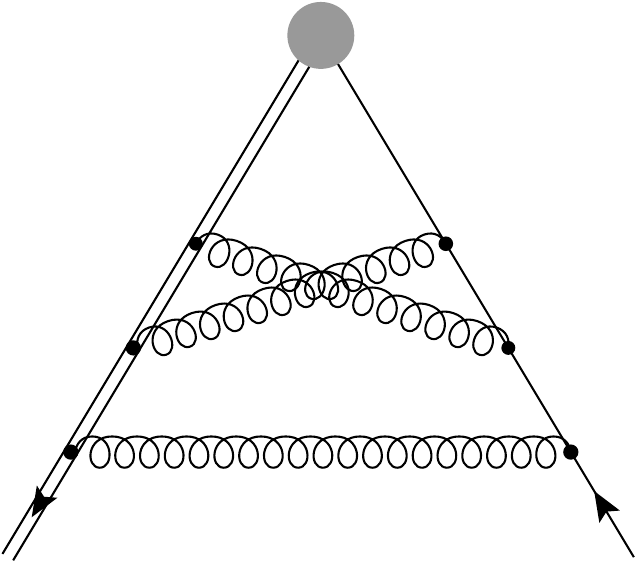}&
			\includegraphics[width=0.2\textwidth]{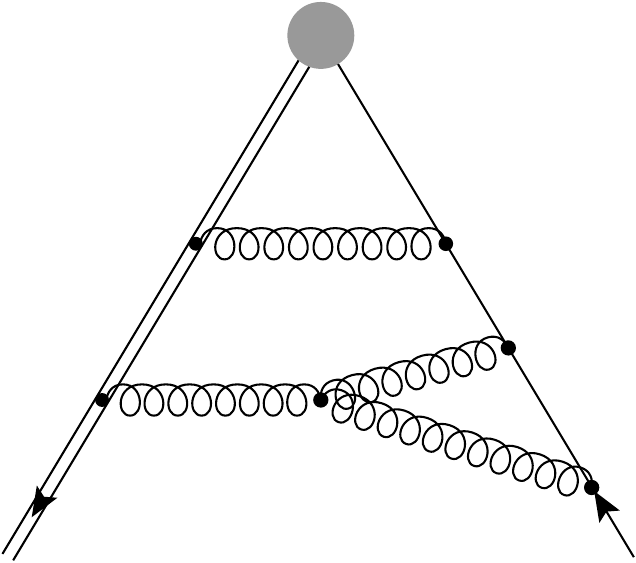}
			\\
			\includegraphics[width=0.2\textwidth]{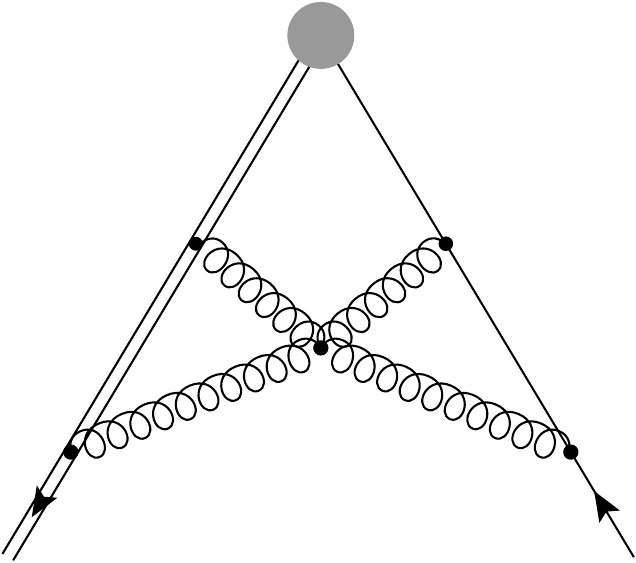}&
			\includegraphics[width=0.2\textwidth]{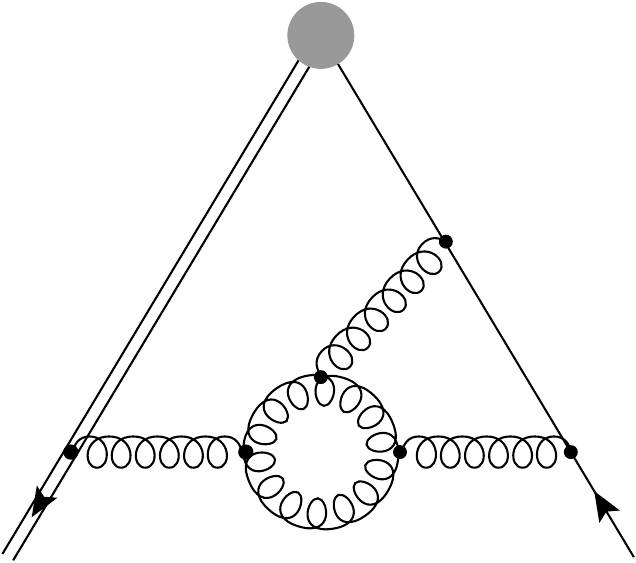}&
			\includegraphics[width=0.2\textwidth]{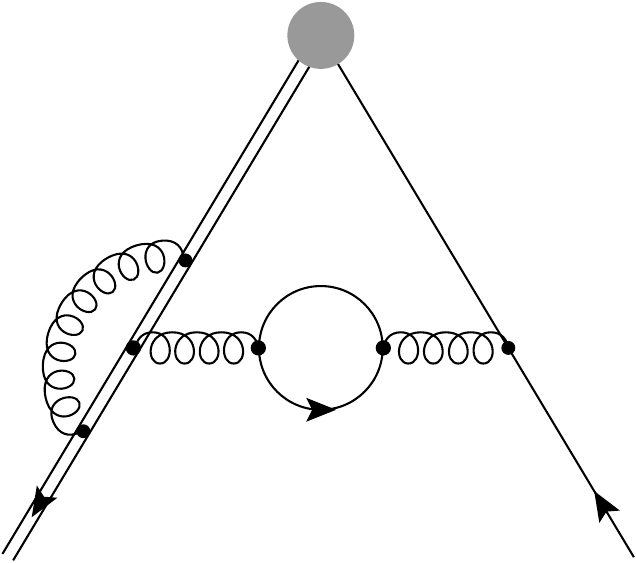}
			\\
			\includegraphics[width=0.2\textwidth]{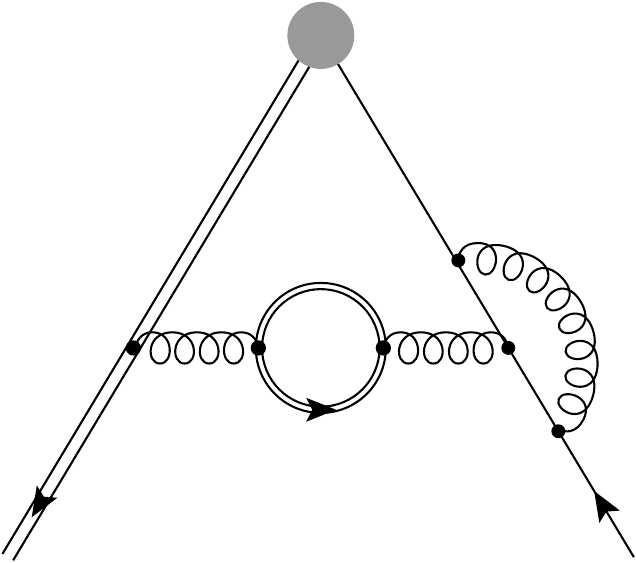}&
			\includegraphics[width=0.2\textwidth]{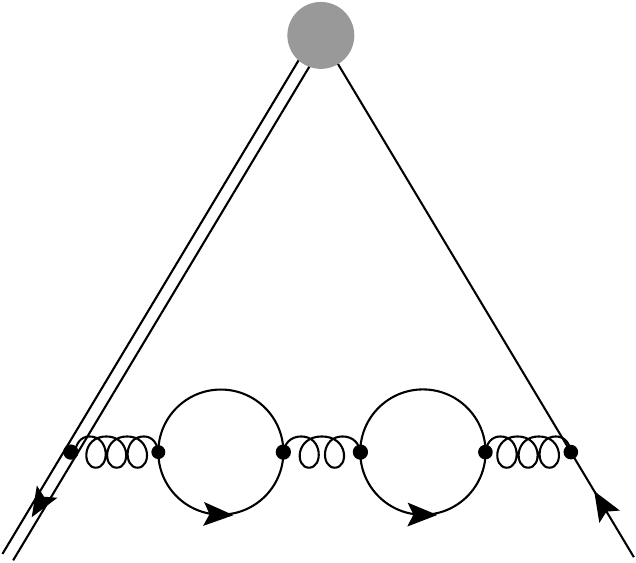}&
			\includegraphics[width=0.2\textwidth]{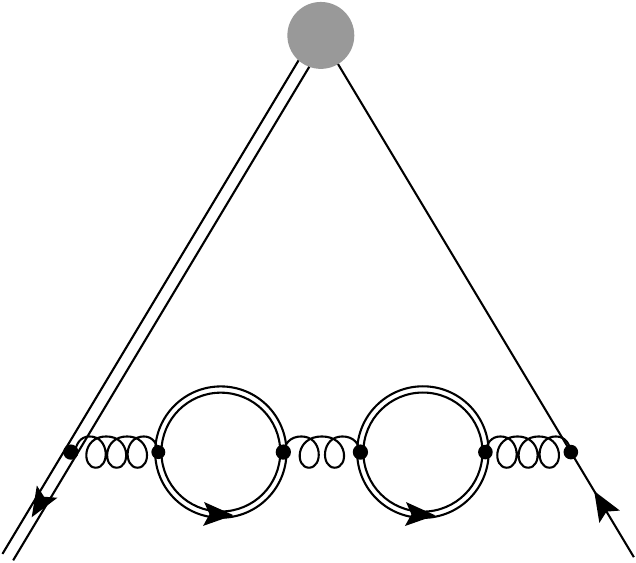}
		\end{tabular}
		\caption{\label{fig::diags}Sample Feynman diagrams contributing to the heavy-to-light form factors. The double solid, solid and curly lines refer to massive quarks, massless quarks and gluons, respectively. The gray blob represents one of the external currents given in Eq.~(\ref{eq::currents}).}
	\end{center}
\end{figure}

In Fig. \ref{fig::diags} we show a set of sample Feynman diagrams for the heavy-to-light form factors.

One of the first steps in our calculation is the application of
projectors for the scalar form factors introduced in
Eq.~(\ref{eq::Gamma}). Explicit expressions are given in
Appendix~\ref{app::proj}.  Afterwards there are no open indices and
all the scalar products can be decomposed into denominator factors
used to define the integral families. For this step we use an
auxiliary file generated by {\tt tapir}.  In total we have
contributions from 47 integral families.  We extract the
respective lists of integrals which serve as input for the integral
reduction. For all external currents we
generate the corresponding amplitude for general QCD gauge parameter $\xi$.

\subsection{Integral reduction}

In a next step, we want to reduce the list of integrals contributing
to the amplitude to a smaller set of master integrals using
integration-by-parts
relations~\cite{Tkachov:1981wb,Chetyrkin:1981qh,Gehrmann:1999as} and
the Laporta algorithm~\cite{Laporta:2000dsw}.  Before performing
the actual reduction for the amplitude, we reduce sample integrals with up
to two dots and one scalar product for each integral family using
\texttt{Kira}~\cite{Maierhofer:2017gsa,Klappert:2020nbg}, employing
\texttt{Fermat}~\cite{Fermat} as computational backend.  These samples
allow us to find a basis of master integrals in which the dependence
on the space-time $d$ and the kinematic variable $s/m^2$ factorizes in
the denominators of all coefficients appearing in the final reduction
tables~\cite{Smirnov:2020quc,Usovitsch:2020jrk}.  We achieve this as
well as an reduction of spurious poles in $\epsilon$ with an improved version
of the code \texttt{ImproveMasters.m}~\cite{Smirnov:2020quc}.

With the basis chosen, we then perform the reductions of all integral
families again employing \texttt{Kira}, this time exploiting the
finite field
techniques~\cite{Kauers:2008zz,vonManteuffel:2014ixa,Peraro:2016wsq}
implemented in
\texttt{FireFly}~\cite{Klappert:2019emp,Klappert:2020aqs}.\footnote{While we managed to complete the reduction after fixing the gauge to $\xi = 0$ with \texttt{Kira 2.3}, we resorted to the current development version to perform the reduction for general $\xi$. We thank Johann Usovitsch and Zihao Wu for allowing us to use the development version (see Ref.~\cite{Driesse:2024xad} for a first brief discussion of some of the improvements).}
In addition to the separate reductions of all families, we run
\texttt{Kira} to find symmetries between the master integrals and
arrive at a set of $429$ master integrals at the three-loop level.  We
then use \texttt{LiteRed}~\cite{Lee:2012cn,Lee:2013mka} and a
subsequent reduction with \texttt{Kira} and \texttt{FireFly} to
establish differential equations for the master
integrals~\cite{Kotikov:1990kg,Kotikov:1991hm,Kotikov:1991pm,Remiddi:1997ny}
in $s/m^2$.

\subsection{\label{sub::MIs}Master integrals}

We calculate the master integrals at one and two loops
analytically.  Additionally, we also consider the master integrals
which contribute to the leading-color amplitude, the ones depending
on the number of light flavors, and the ones with two closed heavy-fermion loops analytically.  The master integrals
contributing to the leading color amplitude have been obtained before
in Refs.~\cite{Chen:2018fwb,Datta:2023otd}.  In the second reference
also the leading color amplitudes for the vector, axialvector, scalar
and pseudoscalar currents have been obtained.  We consider in addition the
tensor current.  For the calculation we use the techniques of
Ref.~\cite{Ablinger:2018zwz}.  In practice this means that we do not
try to find a canonical basis of master integrals, but we uncouple
blocks of the differential equation into higher-order ones and solve
these via the factorization of the differential operator and variation
of constants.  This technique is successful for the considered subset
of master integrals since the differential operators factorize to
first order and the results can therefore be expressed as iterated
integrals over algebraic letters.  We checked explicitly that this is
not the case for the full amplitude, where also elliptic sectors
contribute.  For the implementation of the algorithms we make use of
the packages \texttt{Sigma}~\cite{sigmaI,sigmaII} and
\texttt{HarmonicSums}~\cite{Blumlein:1998if,Vermaseren:1998uu,Blumlein:2009ta,Ablinger:2009ovq,Ablinger:2011te,Ablinger:2012ufz,Ablinger:2013eba,Ablinger:2013cf,Ablinger:2014bra,Ablinger:2014rba,Ablinger:2015gdg,Ablinger:2018cja}.

The boundary constants for the solution are either obtained by direct
integration, Mellin-Barnes techniques, or using \texttt{PSLQ}~\cite{PSLQ} on
numerical results computed with \texttt{AMFLow}~\cite{Liu:2022chg} implementing the auxiliary-mass flow method~\cite{Liu:2017jxz,Liu:2021wks,Liu:2022mfb} at
the point $s=0$.  Many boundary conditions can also be fixed by
regularity conditions in $s/m^2=0$ and $s/m^2=1$.

We find that we can express our analytical results as iterated integrals over the alphabet
\begin{align*}
    \frac{1}{x}&, &
    \frac{1}{1 \pm x}&, &
    \frac{1}{2-x} ~.
\end{align*}

For the remaining master integrals we use the semi-analytic technique developed in
Ref.~\cite{Fael:2021kyg,Fael:2022rgm,Fael:2022miw,Fael:2023zqr}.
The method is based on series expansions around regular and singular points
of the differential equation.
Two neighboring expansions are then numerically matched at a point where both expansions converge.
We use expansions at the points
\begin{align}
    \frac{s}{m^2} &= \{
    -\infty, -60,-40,-30,-20,-15,-10,-8,-7,-6,-5,-4,-3,-2,-1,-1/2,
    \nonumber \\ &
    0 , 1/4 , 1/2, 3/4, 7/8 , 1 \}  ~,
    \label{eq::som2}
\end{align}
where in each case we used 50 expansion terms.
All but the expansions around $s/m^2=1$ and $s/m^2=-\infty$ are regular Taylor expansions.
We used boundary conditions at the regular point $s/m^2=0$ which we obtained
with the help of \texttt{AMFlow} demanding 100 digits precision.


\section{Analytical results}
\label{sec::results}

As mentioned in Section~\ref{sub::MIs}, we have analytic results for all one- and two-loop form
factors up to order $\epsilon^4$ and $\epsilon^2$, respectively. The
computer-readable expressions for all twelve scalar form factors can be downloaded from Ref.~\cite{progdata} for general renormalization scales $\mu$ and $\nu$ and with the option to renormalize the scalar and pseudoscalar current in the $\overline{\rm MS}$ or in the on-shell scheme. We provide both, expressions where only the ultraviolet counterterms have been introduced, and expressions where in addition the infrared poles have been subtracted.  For illustration we show in the following the result
for $C_1^t$ for $\nu^2=\mu^2=m^2$ up to $\mathcal{O}(\epsilon^0)$ which up to two-loop order reads ($x=s/m^2$):
\begin{eqnarray}
  C_1^{t,(0)} &=& 1\,,\nonumber\\
  C_1^{t,(1)} &=& C_F
    \biggl\{
        -\frac{3}{2}
        -\frac{\pi ^2}{48}
        +\frac{(1-2 x) H_1}{2 x}
        -\frac{1}{2} H_{0,1}
        -H_{1,1}
    \biggr\} \,,\nonumber\\
  C_1^{t,(2)} &=& C_A C_F
    \biggl\{
        -\frac{119851}{20736}
        -\frac{\pi ^4 \big(-49+75 x-201 x^2+31 x^3\big)}{1920 (1-x)^3}
        +\frac{(354-829 x) H_1}{216 x}
        \nonumber \\ &&
        +\zeta_3
        \biggl(
                \frac{398-978 x+681 x^2-137 x^3}{576 (1-x)^3}
                +\frac{7}{8} H_1
        \biggr)
        +\pi ^2
        \biggl(
                \frac{253-830 x-1007 x^2}{3456 (1-x)^2}
                \nonumber \\ &&
                -\frac{\big(
                        24-x+56 x^2+65 x^3\big) H_1}{288 (1-x)^2 x}
                -\frac{(2-x) \big(
                        2-5 x-x^2\big) H_2}{96 (1-x)^3}
                -\frac{(1+x) (1-3 x) H_{-1}}{24 (1-x) x}
                \nonumber \\ &&
                +\frac{\big(
                        1-15 x-6 x^2-4 x^3\big) H_{0,1}}{48 (1-x)^3}
                +\frac{1}{24} H_{0,-1}
                +\frac{1}{12} H_{1,1}
                \nonumber \\ &&
                -\frac{\big(
                        32-48 x+9 x^2+11 x^3\big) \ln(2)}{96 (1-x)^3}
        \biggr)
        +\frac{\big(
                66-308 x+259 x^2-149 x^3\big) H_{0,1}}{144 (1-x)^2 x}
        \nonumber \\ &&
        +\frac{(2-5 x) (39-83 x) H_{1,1}}{144 (1-x) x}
        +\frac{\big(
                8+12 x+9 x^2-25 x^3\big) H_{0,0,1}}{48 (1-x)^3}
        \nonumber \\ &&
        -\frac{\big(
                20-192 x+237 x^2-61 x^3\big) H_{0,1,1}}{48 (1-x)^3}
        -\frac{\big(
                15-16 x+17 x^2\big) H_{1,0,1}}{16 (1-x)^2}
        -\frac{11}{6} H_{1,1,1}
        \nonumber \\ &&
        -\frac{(2-x) \big(
                2-5 x-x^2\big) H_{2,1,1}}{48 (1-x)^3}
        -\frac{(1+x) (1-3 x) H_{-1,0,1}}{2 (1-x) x}
        +
        \frac{\big(
                1+x+6 x^2\big) H_{0,0,0,1}}{8 (1-x)^3}
        \nonumber \\ &&
        +\frac{\big(
                1+x+6 x^2\big) H_{0,0,1,1}}{4 (1-x)^3}
        -\frac{\big(
                1+x+6 x^2\big) H_{0,1,0,1}}{8 (1-x)^3}
        +\frac{1}{2} H_{0,-1,0,1}
        +\frac{1}{2} H_{1,0,0,1}
    \biggr\}
    \nonumber \\ &&
    +C_F^2
    \biggl\{
        \frac{2515}{768}
        +\frac{\pi ^4 \big(
                389+7473 x-1857 x^2+907 x^3\big)}{23040 (1-x)^3}
        +H_{1,0,1,1}
        -\frac{(3-8 x) H_1}{4 x}
        \nonumber \\ &&
        -\zeta_3
        \biggl(
                 \frac{26-54 x+21 x^2+3 x^3}{32 (1-x)^3}
                +\frac{1}{2} H_1
        \biggr)
        +\pi ^2
        \biggl(
                -\frac{51+47 x+2 x^2}{48 (1-x)^2}
                \nonumber \\ &&
                +\frac{\big(
                        11-114 x-105 x^2+16 x^3\big) H_1}{96 (1-x)^2 x}
                +\frac{(2-x) \big(
                        2-5 x-x^2\big) H_2}{48 (1-x)^3}
                \nonumber \\ &&
                +\frac{(1+x) (1-3 x) H_{-1}}{12 (1-x) x}
                -\frac{\big(
                        35+135 x+21 x^2+x^3\big) H_{0,1}}{96 (1-x)^3}
                -\frac{1}{12} H_{0,-1}
                +\frac{1}{48} H_{1,1}
                \nonumber \\ &&
                +\frac{\big(
                        32-48 x+9 x^2+11 x^3\big) \ln(2)}{48 (1-x)^3}
        \biggr)
        -\frac{\big(
                6-42 x+91 x^2+45 x^3\big) H_{0,1}}{24 (1-x)^2 x}
        \nonumber \\ &&
        +\frac{(97+3 x) H_{1,1}}{24 (1-x)}
        +\frac{\big(
                6+12 x+84 x^2-141 x^3+35 x^4\big) H_{0,0,1}}{24 (1-x)^3 x}
                -\frac{3 (1-2 x) H_{1,1,1}}{2 x}
        \nonumber \\ &&
        -\frac{\big(
                12-150 x+126 x^2-51 x^3+59 x^4\big) H_{0,1,1}
        }{24 (1-x)^3 x}
        +
        \frac{\big(
                -41-68 x+13 x^2\big) H_{1,0,1}}{24 (-1+x)^2}
        \nonumber \\ &&
        -\frac{(-2+x) \big(
                -2+5 x+x^2\big) H_{2,1,1}}{24 (-1+x)^3}
        +\frac{(1+x) (-1+3 x) H_{-1,0,1}}{(-1+x) x}
        \nonumber \\ &&
        -\frac{\big(
                1+17 x-4 x^2+2 x^3\big) H_{0,0,0,1}}{4 (-1+x)^3}
        -\frac{\big(
                3+11 x+2 x^2\big) H_{0,0,1,1}}{2 (-1+x)^3}
        +\frac{1}{2} H_{1,1,0,1}
        +3 H_{1,1,1,1}
        \nonumber \\ &&
        +\frac{\big(
                2+14 x-x^2+x^3\big) H_{0,1,0,1}}{4 (-1+x)^3}
        +\frac{3}{2} H_{0,1,1,1}
        -H_{0,-1,0,1}
        -H_{1,0,0,1}
    \biggr\}
    \nonumber \\ &&
    + C_F T_F n_h
    \biggl\{
        \frac{2267-5398 x+4283 x^2}{1296 (-1+x)^2}
        -\frac{\pi ^2 \big(
                -11+45 x-57 x^2+7 x^3\big)}{108 (-1+x)^3}
        \nonumber \\ &&
        +\frac{2 \big(
                -6+13 x-14 x^2+19 x^3\big) H_1}{27 (-1+x)^2 x}
        -\frac{(1+x) \big(
                -3+5 x-5 x^2+11 x^3\big) H_{0,1}}{18 (-1+x)^3 x}
        \nonumber \\ &&
        +\frac{1}{6} H_{0,0,1}
        -\frac{\zeta_{3}}{6}
    \biggr\}
    + C_F T_F n_l
    \biggl\{
        \frac{7859}{5184}
        +\frac{1}{864} \pi ^2 \big(
                109+48 H_1\big)
        +\frac{(-12+31 x) H_1}{27 x}
        \nonumber \\ &&
        -\frac{(3-11 x) H_{0,1}}{18 x}
        -\frac{(3-11 x) H_{1,1}}{9 x}
        +\frac{1}{6} H_{0,0,1}
        +\frac{1}{3} H_{0,1,1}
        +\frac{1}{3} H_{1,0,1}
        +\frac{2}{3} H_{1,1,1}
        \nonumber \\ &&
        +\frac{13 \zeta_3}{72}
    \biggr\}\,.
\end{eqnarray}
The one-loop order has successfully been compared to Ref.~\cite{Bell:2010mg}
up to order $\epsilon^2$ and has been extended to $\epsilon^4$.
Similarly, our two-loop results up to the constant part in $\epsilon$
agrees with Ref.~\cite{Bell:2010mg} and we have added $\epsilon^1$
and $\epsilon^2$ terms.

After multiplying $\Gamma_{\mu\nu}^t(q_1,q_2)$ in Eq.~(\ref{eq::Gamma}) with $q^\nu$
and projecting the result to $F_2^v$ we obtain the
contribution for $b\to s \gamma$ which is given by
\begin{eqnarray}
  F_1^{t} - \frac{1}{2} F_2^{t} - \frac{1}{2} F_3^{t}
  \,.
\end{eqnarray}
Using our analytic results we
find agreement with the numerical expressions given in Eqs.~(88) and~(89) of Ref.~\cite{Ali:2007sj}.

At three-loop order the amplitude can be
divided up into the different color factors\footnote{The same color decomposition also holds for the infrared subtracted quantities $C$.}
\begin{eqnarray}
    F_{i}^{x,(3)} &=&
      C_F T_F^2 n_l^2 F_{i}^{x,(3),n_l^2}
    + C_F T_F^2 n_h^2 F_{i}^{x,(3),n_h^2}
    + C_F T_F^2 n_l n_h F_{i}^{x,(3),n_l n_h}
    \nonumber \\ &&
    + C_F^2 T_F n_l F_{i}^{x,(3),C_F n_l}
    + C_F C_A T_F n_l F_{i}^{x,(3),C_A n_l}
    + N_C^3 F_{i}^{x,(3), N_C^3}
    \nonumber \\ &&
    + C_F^2 n_h T_F F_{i}^{x,(3),C_F n_h}
    + C_F C_A n_h T_F F_{i}^{x,(3),C_A n_h}
    + \mathcal{O}\left(N_C^2\right)
\end{eqnarray}
up to color suppressed contributions.
We have computed the first six terms analytically.
The corresponding expressions can again be downloaded from
the webpage~\cite{progdata}. The explicit three-loop expressions for the tensor coefficient $C_1^t$ read
\begin{eqnarray}
  C_1^{t,(3),n_l^2} &=& -\frac{370949}{419904}
-\frac{221 \pi ^4}{38880}
-\pi ^2 \biggl(
         \frac{829}{3888}
        -\frac{(3-11 x) H_1}{81 x}
        +\frac{1}{27} H_{0,1}
        +\frac{2}{27} H_{1,1}
\biggr)
\nonumber \\ &&
+\frac{(657-1430 x) H_1}{1458 x}
+\frac{(48-121 x) H_{0,1}}{162 x}
+\frac{(48-121 x) H_{1,1}}{81 x}
\nonumber \\ &&
+\frac{(3-11 x) H_{0,0,1}}{27 x}
+\frac{2 (3-11 x) H_{0,1,1}}{27 x}
+\frac{2 (3-11 x) H_{1,0,1}}{27 x}
-\frac{4}{9} H_{1,1,0,1}
\nonumber \\ &&
+\frac{4 (3-11 x) H_{1,1,1}}{27 x}
-\frac{1}{9} H_{0,0,0,1}
-\frac{2}{9} H_{0,0,1,1}
-\frac{2}{9} H_{0,1,0,1}
-\frac{4}{9} H_{0,1,1,1}
\nonumber \\ &&
-\frac{2}{9} H_{1,0,0,1}
-\frac{4}{9} H_{1,0,1,1}
-\frac{8}{9} H_{1,1,1,1}
-\frac{\big(
        323+126 H_1\big) \zeta_3}{486}
~, \\
  C_1^{t,(3),n_h^2} &=&
    -\frac{\pi ^4}{540}
    +\frac{\pi ^2 (5+3 x)}{135 (1-x)}
    -\frac{667-2704 x+4273 x^2-3070 x^3+810 x^4}{324 (1-x)^4}
    \nonumber \\ &&
    +\zeta_{3}
    \biggl(
        \frac{21-162 x+483 x^2-668 x^3+477 x^4-198 x^5+39 x^6}{27 (-1+x)^6}
        +\frac{2}{9} H_1
    \biggr)
    \nonumber \\ &&
    +\frac{\big(
        438-2147 x+4124 x^2-4926 x^3+3734 x^4-1079 x^5\big) H_1}{972 (1-x)^4 x}
    \nonumber \\ &&
    +\frac{(1+x) \big(
        48-193 x+282 x^2-346 x^3+354 x^4-121 x^5\big) H_{0,1}}{162 (1-x)^5 x}
    \nonumber \\ &&
    +\frac{\big(
        3-11 x+39 x^2-123 x^3+153 x^4-75 x^5+33 x^6-11 x^7\big)}{27 (1-x)^6 x}  H_{0,0,1}
    \nonumber \\ &&
    -\frac{1}{9} H_{0,0,0,1}
    -\frac{2}{9} H_{1,0,0,1}
~, \\
  C_1^{t,(3),n_h n_l} &=&
  -\frac{\pi ^4}{810}
-\frac{23611-43766 x+9787 x^2}{5832 (1-x)^2}
+\pi ^2
\biggl(
        \frac{371-585 x-327 x^2+349 x^3}{1944 (1-x)^3}
        \nonumber \\ &&
        -\frac{(1+x) \big(
                3-5 x+5 x^2-11 x^3\big) H_1}{81 (1-x)^3 x}
        +\frac{1}{27} H_{0,1}
\biggr)
\nonumber \\ &&
+\frac{\big(
        73-308 x+421 x^2-138 x^3\big) H_1}{81 (1-x)^2 x}
-\frac{2}{9} H_{0,0,0,1}
-\frac{2}{9} H_{0,0,1,1}
\nonumber \\ &&
+\frac{\big(
        48-157 x+153 x^2-27 x^3+31 x^4\big) H_{0,1}}{81 (1-x)^3 x}
\nonumber \\ &&
+\frac{8 \big(
        6-13 x+14 x^2-19 x^3\big) H_{1,1}}{81 (1-x)^2 x}
+\frac{2 \big(
        3-11 x+27 x^2-24 x^3-11 x^4\big) H_{0,0,1}}{27 (1-x)^3 x}
\nonumber \\ &&
+\frac{2 (1+x) \big(
        3-5 x+5 x^2-11 x^3\big) H_{0,1,1}}{27 (1-x)^3 x}
-\frac{2}{9} H_{0,1,0,1}
+\frac{2 \big(
        10-20 x+x^2\big) \zeta_{3}}{27 (1-x)^2}
\nonumber \\ &&
+\frac{2 (1+x) \big(
        3-5 x+5 x^2-11 x^3\big) H_{1,0,1}}{27 (1-x)^3 x}
~, \\
  C_1^{t,(3),C_R n_l} &=&
  -\frac{82223}{62208}+H_{0,-1,0,1,1}
+\pi ^4 \biggl(
        \frac{52459+113919 x-71799 x^2+48173 x^3}{622080 (-1+x)^3}
\nonumber \\ &&
        +\frac{61 H_1}{4320}
\biggr)
+\zeta_3
\biggl(
        -\frac{6142-6438 x-3345 x^2+3155 x^3}{2592 (-1+x)^3}
        \nonumber \\ &&
        -\frac{\big(
                -73+70 x-473 x^2+284 x^3\big) H_1}{144 (-1+x)^2 x}
        +\frac{7 (-2+x) \big(
                -2+5 x+x^2\big) H_2}{144 (-1+x)^3}
                \nonumber \\ &&
        -\frac{(1+x) (-1+3 x) H_{-1}}{4 (-1+x) x}
        -\frac{\big(
                -49+387 x-231 x^2+85 x^3\big) H_{0,1}}{144 (-1+x)^3}
        +\frac{1}{4} H_{0,-1}
        \nonumber \\ &&
        -\frac{91}{72} H_{1,1}
\biggr)
+\pi ^2
\biggl[
        \frac{575209+500974 x+5545 x^2}{248832 (-1+x)^2}
        \nonumber \\ &&
        +\ln(2) \biggl(
                \frac{280-516 x+249 x^2+5 x^3}{144 (-1+x)^3}
                +\frac{(7+17 x) H_1}{36 (-1+x)}
                -\frac{1}{6} H_{0,-1}
                +\frac{1}{6} H_{1,1}
                \nonumber \\ &&
                +\frac{(-2+x) \big(
                        -2+5 x+x^2\big) H_2}{72 (-1+x)^3}
                +\frac{(1+x) (-1+3 x) H_{-1}}{6 (-1+x) x}
        \biggr)
        \nonumber \\ &&
        -\frac{\big(
                1569-14528 x-11363 x^2+2434 x^3\big) H_1}{5184 (-1+x)^2 x}
        +\frac{(2-x) \big(
                26-31 x+23 x^2\big) H_2}{144 (1-x)^3}
        \nonumber \\ &&
        -\frac{(1+x) \big(
                41-131 x+102 x^2\big) H_{-1}}{216 (-1+x)^2 x}
        \nonumber \\ &&
        -\frac{\big(
                -402+2899 x+1347 x^2-2175 x^3+923 x^4\big) H_{0,1}
        }{1728 (-1+x)^3 x}
        \nonumber \\ &&
        -
        \frac{\big(
                -12+71 x-117 x^2+15 x^3+19 x^4\big) H_{0,-1}}{216 (-1+x)^3 x}
        \nonumber \\ &&
        -\frac{\big(
                90-1117 x-1636 x^2+359 x^3\big) H_{1,1}}{864 (-1+x)^2 x}
        +\frac{(6+x) H_{1,2}}{36 x}
        \nonumber \\ &&
        -\frac{(1+x) (-1+3 x) H_{1,-1}}{18 (-1+x) x}
        +\frac{(-2+x) \big(
                -2+5 x+x^2\big) H_{2,1}}{72 (-1+x)^3}
        \nonumber \\ &&
        -\frac{(-2+x) \big(
                -2+5 x+x^2\big) H_{2,2}}{72 (-1+x)^3}
        +\frac{(1+x) (-1+3 x) H_{-1,1}}{9 (-1+x) x}
        \nonumber \\ &&
        +\frac{(1+x) (-1+3 x) H_{-1,-1}}{18 (-1+x) x}
        +\frac{\big(
                -107-79 x-181 x^2+47 x^3\big) H_{0,0,1}}{288 (-1+x)^3}
        \nonumber \\ &&
        +\frac{1}{18} H_{0,0,-1}
        -\frac{\big(
                21+97 x+7 x^2+3 x^3\big) H_{0,1,1}}{48 (-1+x)^3}
        -\frac{1}{6} H_{0,1,2}
        +\frac{1}{18} H_{0,1,-1}
        \nonumber \\ &&
        -\frac{1}{9} H_{0,-1,1}
        -\frac{1}{18} H_{0,-1,-1}
        -\frac{1}{144} H_{1,0,1}
        -\frac{13}{72} H_{1,1,1}
        -\frac{1}{6} H_{1,1,2}
        \nonumber \\ &&
        -\frac{\big(
                40-72 x+15 x^2+13 x^3\big) \ln^2(2)}{108 (-1+x)^3}
        \nonumber \\ &&
        +\frac{\big(
                725+14145 x-3537 x^2+1723 x^3\big) \zeta_3}{3456 (-1+x)^3}
\biggr]
\nonumber \\ &&
-\frac{(-7103+22516 x) H_1}{10368 x}
+\frac{\big(
        4608-27263 x+80878 x^2+36529 x^3\big) H_{0,1}}{10368 (-1+x)^2 x}
        \nonumber \\ &&
+\frac{\big(
        -432+2976 x+44615 x^2+217 x^3\big) H_{1,1}
}{5184 (-1+x) x^2}
\nonumber \\ &&
+
\frac{\big(
        14+217 x+207 x^2-730 x^3+274 x^4\big) H_{0,0,1}}{72 (-1+x)^3 x}
\nonumber \\ &&
-\frac{\big(
        246-3392 x+3372 x^2-363 x^3+83 x^4\big) H_{0,1,1}}{216 (-1+x)^3 x}
        \nonumber \\ &&
+\frac{\big(
        -90+701 x+1220 x^2+89 x^3\big) H_{1,0,1}}{216 (-1+x)^2 x}
        \nonumber \\ &&
-\frac{\big(
        18-87 x+19 x^2\big) H_{1,1,1}}{9 (-1+x) x}
+\frac{(-2+x) \big(
        26-31 x+23 x^2\big) H_{2,1,1}}{72 (-1+x)^3}
        \nonumber \\ &&
-\frac{(1+x) \big(
        41-131 x+102 x^2\big) H_{-1,0,1}}{18 (-1+x)^2 x}
        \nonumber \\ &&
+\frac{\big(
        12+22 x+338 x^2-415 x^3+127 x^4\big) H_{0,0,0,1}}{36 (-1+x)^3 x}
        \nonumber \\ &&
-\frac{\big(
        12-285 x-121 x^2+128 x^3+22 x^4\big) H_{0,0,1,1}}{36 (-1+x)^3 x}
        \nonumber \\ &&
-\frac{\big(
        -9+34 x+13 x^2+25 x^3+x^4\big) H_{0,1,0,1}}{18 (-1+x)^3 x}
        \nonumber \\ &&
-\frac{\big(
        21-231 x+339 x^2-270 x^3+137 x^4\big) H_{0,1,1,1}}{18 (-1+x)^3 x}
        \nonumber \\ &&
-\frac{\big(
        -12+71 x-117 x^2+15 x^3+19 x^4\big) H_{0,-1,0,1}}{18 (-1+x)^3 x}
        \nonumber \\ &&
+\frac{\big(
        -24+17 x-238 x^2+101 x^3\big) H_{1,0,0,1}}{36 (-1+x)^2 x}
        \nonumber \\ &&
-\frac{\big(
        -15+140 x-109 x^2+80 x^3\big) H_{1,0,1,1}}{18 (-1+x)^2 x}
        \nonumber \\ &&
-\frac{\big(
        -12-5 x-254 x^2+79 x^3\big) H_{1,1,0,1}}{36 (-1+x)^2 x}
-\frac{10 (-1+3 x) H_{1,1,1,1}
}{3 x}
\nonumber \\ &&
+\frac{(6+x) H_{1,2,1,1}}{18 x}
-\frac{2 (1+x) (-1+3 x) H_{1,-1,0,1}}{3 (-1+x) x}
\nonumber \\ &&
+\frac{(-2+x) \big(
        -2+5 x+x^2\big) H_{2,1,1,1}}{18 (-1+x)^3}
-\frac{(-2+x) \big(
        -2+5 x+x^2\big) H_{2,2,1,1}}{36 (-1+x)^3}
\nonumber \\ &&
-\frac{5 (1+x) (-1+3 x) H_{-1,0,0,1}}{6 (-1+x) x}
-\frac{(1+x) (-1+3 x) H_{-1,0,1,1}}{(-1+x) x}
\nonumber \\ &&
-\frac{2 (1+x) (-1+3 x) H_{-1,1,0,1}}{3 (-1+x) x}
+\frac{2 (1+x) (-1+3 x) H_{-1,-1,0,1}}{3 (-1+x) x}
\nonumber \\ &&
+\frac{\big(
        1+17 x-4 x^2+2 x^3\big) H_{0,0,0,0,1}}{3 (-1+x)^3}
+\frac{\big(
        5+25 x+x^2+x^3\big) H_{0,0,0,1,1}}{3 (-1+x)^3}
        \nonumber \\ &&
+\frac{\big(
        -4+32 x-19 x^2+7 x^3\big) H_{0,0,1,0,1}}{12 (-1+x)^3}
-\frac{(1+x) \big(
        15+20 x-3 x^2\big) H_{0,0,1,1,1}}{6 (1-x)^3}
        \nonumber \\ &&
+\frac{2}{3} H_{0,0,-1,0,1}
+\frac{\big(
        -2+66 x-27 x^2+11 x^3\big) H_{0,1,0,0,1}}{12 (-1+x)^3}
        \nonumber \\ &&
-\frac{\big(
        -9-13 x-13 x^2+3 x^3\big) H_{0,1,0,1,1}}{6 (-1+x)^3}
-\frac{\big(
        1+17 x-4 x^2+2 x^3\big) H_{0,1,1,0,1}}{3 (-1+x)^3}
        \nonumber \\ &&
-\frac{10}{3} H_{0,1,1,1,1}
-\frac{1}{3} H_{0,1,2,1,1}
+\frac{2}{3} H_{0,1,-1,0,1}
+\frac{5}{6} H_{0,-1,0,0,1}
+\frac{2}{3} H_{0,-1,1,0,1}
\nonumber \\ &&
-\frac{2}{3} H_{0,-1,-1,0,1}
+2 H_{1,0,0,0,1}
+\frac{2}{3} H_{1,0,1,0,1}
-3 H_{1,0,1,1,1}
+\frac{11}{6} H_{1,1,0,0,1}
\nonumber \\ &&
-2 H_{1,1,0,1,1}
-\frac{4}{3} H_{1,1,1,0,1}
-\frac{20}{3} H_{1,1,1,1,1}
-\frac{1}{3} H_{1,1,2,1,1}
\nonumber \\ &&
-\frac{\big(
        68-108 x+21 x^2+23 x^3\big) \ln^4(2)}{432 (-1+x)^3}
-\frac{\big(
        68-108 x+21 x^2+23 x^3\big) \mathrm{Li}_4\!\left(\frac{1}{2}\right)}{18 (-1+x)^3}
\nonumber \\ &&
+\frac{\big(
        -13-21 x-18 x^2+4 x^3\big) \zeta_{5}}{9 (-1+x)^3}
~, \\
  C_1^{t,(3),C_A n_l} &=&
\frac{4126157}{419904}
+\pi ^4 \biggl(
        -\frac{113-9573 x-7707 x^2+2935 x^3}{155520 (-1+x)^3}
        -\frac{19}{864} H_1
\biggr)
\nonumber \\ &&
+\zeta_3 \biggl(
        \frac{-5092+21000 x-29703 x^2+13309 x^3}{5184 (-1+x)^3}
        -\frac{1}{8} H_{0,-1}
        +\frac{3}{8} H_{1,1}
        \nonumber \\ &&
        +\frac{\big(
                -36+113 x-202 x^2+173 x^3\big) H_1}{144 (-1+x)^2 x}
        -\frac{7 (-2+x) \big(
                -2+5 x+x^2\big) H_2}{288 (-1+x)^3}
        \nonumber \\ &&
        +\frac{(1+x) (-1+3 x) H_{-1}}{8 (-1+x) x}
        +\frac{\big(
                -7+17 x-24 x^2+6 x^3\big) H_{0,1}}{24 (-1+x)^3}
\biggr)
\nonumber \\ &&
+\pi ^2 \biggl[
        \frac{236191-273122 x+373243 x^2}{279936 (-1+x)^2}
        +\ln(2) \biggl(
                -\frac{280-516 x+249 x^2+5 x^3}{288 (-1+x)^3}
                \nonumber \\ &&
                -\frac{(7+17 x) H_1}{72 (-1+x)}
                -\frac{(-2+x) \big(
                        -2+5 x+x^2\big) H_2}{144 (-1+x)^3}
                -\frac{(1+x) (-1+3 x) H_{-1}}{12 (-1+x) x}
                \nonumber \\ &&
                +\frac{1}{12} H_{0,-1}
                -\frac{1}{12} H_{1,1}
        \biggr)
        +\frac{\big(
                -108+8077 x-5921 x^2+7780 x^3\big) H_1}{7776 (-1+x)^2 x}
        \nonumber \\ &&
        -\frac{(-2+x) \big(
                26-31 x+23 x^2\big) H_2}{288 (-1+x)^3}
        +\frac{(1+x) \big(
                41-131 x+102 x^2\big) H_{-1}}{432 (-1+x)^2 x}
        \nonumber \\ &&
        +\frac{\big(
                -16+4 x-110 x^2-145 x^3+87 x^4\big) H_{0,1}
        }{288 (-1+x)^3 x}
        \nonumber \\ &&
        +
        \frac{\big(
                -12+71 x-117 x^2+15 x^3+19 x^4\big) H_{0,-1}}{432 (-1+x)^3 x}
        \nonumber \\ &&
        +\frac{\big(
                96+71 x+74 x^2+335 x^3\big) H_{1,1}}{864 (-1+x)^2 x}
        -\frac{(6+x) H_{1,2}}{72 x}
        +\frac{(1+x) (-1+3 x) H_{1,-1}}{36 (-1+x) x}
        \nonumber \\ &&
        -\frac{(-2+x) \big(
                -2+5 x+x^2\big) H_{2,1}}{144 (-1+x)^3}
        +\frac{(-2+x) \big(
                -2+5 x+x^2\big) H_{2,2}}{144 (-1+x)^3}
        \nonumber \\ &&
        -\frac{(1+x) (-1+3 x) H_{-1,1}}{18 (-1+x) x}
        -\frac{(1+x) (-1+3 x) H_{-1,-1}}{36 (-1+x) x}
        \nonumber \\ &&
        -\frac{\big(
                -3+29 x+6 x^2+8 x^3\big) H_{0,0,1}}{144 (-1+x)^3}
        -\frac{1}{36} H_{0,0,-1}
        +\frac{1}{12} H_{0,1,2}
        -\frac{1}{36} H_{0,1,-1}
        \nonumber \\ &&
        -\frac{\big(
                -1+15 x+6 x^2+4 x^3\big) H_{0,1,1}}{36 (-1+x)^3}
        +\frac{1}{36} H_{0,-1,-1}
        +\frac{1}{18} H_{0,-1,1}
        -\frac{1}{9} H_{1,1,1}
        \nonumber \\ &&
        +\frac{1}{12} H_{1,1,2}
        -\frac{\big(
                40-72 x+15 x^2+13 x^3\big) \ln^2(2)}{216 (1-x)^3}
        \nonumber \\ &&
        -\frac{\big(
                53-23 x+261 x^2-19 x^3\big) \zeta_3}{288 (1-x)^3}
\biggr]
+\frac{(-98586+201431 x) H_1}{23328 x}
\nonumber \\ &&
+\frac{\big(
        -2742+10661 x-10999 x^2+5906 x^3\big) H_{0,1}}{1296 (-1+x)^2 x}
        \nonumber \\ &&
+\frac{\big(
        4431-14722 x+13117 x^2\big) H_{1,1}
}{1296 (-1+x) x}
\nonumber \\ &&
+
\frac{\big(
        264-968 x+2619 x^2-2079 x^3+218 x^4\big) H_{0,0,1}}{432 (-1+x)^3 x}
\nonumber \\ &&
+\frac{\big(
        420-2539 x+7119 x^2-7008 x^3+1954 x^4\big) H_{0,1,1}}{432 (-1+x)^3 x}
\nonumber \\ &&
+\frac{\big(
        -501+3274 x-3920 x^2+2143 x^3\big) H_{1,0,1}}{432 (-1+x)^2 x}
\nonumber \\ &&
+\frac{\big(
        393-2044 x+1915 x^2\big) H_{1,1,1}}{216 (-1+x) x}
-\frac{(-2+x) \big(
        26-31 x+23 x^2\big) H_{2,1,1}}{144 (-1+x)^3}
\nonumber \\ &&
+\frac{(1+x) \big(
        41-131 x+102 x^2\big) H_{-1,0,1}}{36 (-1+x)^2 x}
\nonumber \\ &&
-\frac{\big(
        40+170 x+177 x^2-159 x^3\big) H_{0,0,0,1}}{144 (1-x)^3}
+\frac{\big(
        2+163 x-99 x^2+10 x^3\big) H_{0,0,1,1}}{36 (-1+x)^3}
\nonumber \\ &&
+\frac{\big(
        -145+325 x-414 x^2+94 x^3\big) H_{0,1,0,1}}{144 (-1+x)^3}
\nonumber \\ &&
+\frac{\big(
        -125+651 x-753 x^2+219 x^3\big) H_{0,1,1,1}}{72 (-1+x)^3}
\nonumber \\ &&
+\frac{\big(
        -12+71 x-117 x^2+15 x^3+19 x^4\big) H_{0,-1,0,1}}{36 (-1+x)^3 x}
\nonumber \\ &&
-\frac{\big(
        -3+x+10 x^2+10 x^3\big) H_{1,0,0,1}}{18 (-1+x)^2 x}
+\frac{\big(
        38-109 x+47 x^2\big) H_{1,0,1,1}}{18 (-1+x)^2}
\nonumber \\ &&
+\frac{\big(
        377-586 x+401 x^2\big) H_{1,1,0,1}}{144 (-1+x)^2}
+\frac{44}{9} H_{1,1,1,1}
-\frac{(6+x) H_{1,2,1,1}
}{36 x}
\nonumber \\ &&
+
\frac{(1+x) (-1+3 x) H_{1,-1,0,1}}{3 (-1+x) x}
-\frac{(-2+x) \big(
        -2+5 x+x^2\big) H_{2,1,1,1}}{36 (-1+x)^3}
\nonumber \\ &&
+\frac{(-2+x) \big(
        -2+5 x+x^2\big) H_{2,2,1,1}}{72 (-1+x)^3}
+\frac{5 (1+x) (-1+3 x) H_{-1,0,0,1}}{12 (-1+x) x}
\nonumber \\ &&
+\frac{(1+x) (-1+3 x) H_{-1,0,1,1}}{2 (-1+x) x}
+\frac{(1+x) (-1+3 x) H_{-1,1,0,1}}{3 (-1+x) x}
\nonumber \\ &&
-\frac{(1+x) (-1+3 x) H_{-1,-1,0,1}}{3 (-1+x) x}
+\frac{\big(
        1+x+6 x^2\big) H_{0,0,0,0,1}}{6 (-1+x)^3}
        \nonumber \\ &&
+\frac{\big(
        1+x+6 x^2\big) H_{0,0,0,1,1}}{3 (-1+x)^3}
+\frac{\big(
        1+x+6 x^2\big) H_{0,0,1,0,1}}{24 (-1+x)^3}
+\frac{\big(
        1+x+6 x^2\big) H_{0,0,1,1,1}}{3 (-1+x)^3}
    \nonumber \\ &&
-\frac{1}{3} H_{0,0,-1,0,1}
-\frac{\big(
        -7+9 x-30 x^2+4 x^3\big) H_{0,1,0,0,1}}{24 (-1+x)^3}
+\frac{\big(
        1+x+6 x^2\big) H_{0,1,0,1,1}}{6 (-1+x)^3}
\nonumber \\ &&
-\frac{\big(
        1+x+6 x^2\big) H_{0,1,1,0,1}}{6 (-1+x)^3}
+\frac{1}{6} H_{0,1,2,1,1}
-\frac{1}{3} H_{0,1,-1,0,1}
-\frac{5}{12} H_{0,-1,0,0,1}
\nonumber \\ &&
-\frac{1}{2} H_{0,-1,0,1,1}
-\frac{1}{3} H_{0,-1,1,0,1}
+\frac{1}{3} H_{0,-1,-1,0,1}
-\frac{11}{12} H_{1,0,0,0,1}
-\frac{2}{3} H_{1,0,0,1,1}
\nonumber \\ &&
-\frac{1}{3} H_{1,0,1,0,1}
-H_{1,1,0,0,1}
-\frac{1}{6} H_{1,1,0,1,1}
+\frac{1}{6} H_{1,1,2,1,1}
\nonumber \\ &&
+\frac{\big(
        68-108 x+21 x^2+23 x^3\big) \ln^4(2)}{864 (-1+x)^3}
+\frac{\big(
        68-108 x+21 x^2+23 x^3\big)\mathrm{Li}_4\!\left(\frac{1}{2}\right)}{36 (-1+x)^3}
\nonumber \\ &&
-\frac{\big(
        -43+161 x-105 x^2+51 x^3\big) \zeta_{5}}{48 (-1+x)^3}
~, \\
  C_1^{t,(3),N_c^3} &=&
  -\frac{155263507}{26873856}
  -\frac{\pi ^6 \big(
        -7514867-19812135 x-18106521 x^2+692147 x^3\big)}{1672151040 (-1+x)^3}
 \nonumber \\ &&
+\zeta_{5} \biggl(
        \frac{1199+6609 x+3760 x^2+162 x^3}{768 (-1+x)^3}
        -\frac{\big(
                -18-12 x-43 x^2+7 x^3\big) H_1}{16 (-1+x)^3}
\biggr)
\nonumber \\ &&
+\pi ^4 \biggl(
        -\frac{347993+1668009 x-548373 x^2+39943 x^3}{4976640 (-1+x)^3}
        \nonumber \\ &&
        +\frac{\big(
                555-14839 x+13219 x^2+18643 x^3+8342 x^4\big) H_1}{368640 (-1+x)^3 x}
                \nonumber \\ &&
        -\frac{\big(
                2691+21975 x+7817 x^2+2749 x^3\big) H_{0,1}}{368640 (-1+x)^3}
                \nonumber \\ &&
        -\frac{\big(
                -3053-1209 x-7431 x^2+1325 x^3\big) H_{1,1}}{184320 (-1+x)^3}
\biggr)
\nonumber \\ &&
+\zeta_{3} \biggl(
        \frac{41639-91621 x+51926 x^2}{62208 (-1+x)^2}
        -\frac{\big(
                570-7411 x+3221 x^2-4264 x^3\big) H_1}{3456 (1-x)^2 x}
                \nonumber \\ &&
        +\frac{\big(
                -18-703 x-1779 x^2-2751 x^3+139 x^4\big) H_{0,1}}{1152 (-1+x)^3 x}
                -\frac{15}{32} H_{1,1,1}
                \nonumber \\ &&
        +\frac{\big(
                90+118 x-479 x^2+325 x^3\big) H_{1,1}}{576 (-1+x)^2 x}
        +\frac{(-1+2 x) (1+8 x) H_{0,0,1}}{32 (-1+x)^3}
        \nonumber \\ &&
        -\frac{\big(
                -5+15 x-12 x^2+5 x^3\big) H_{0,1,1}}{32 (-1+x)^3}
        -\frac{\big(
                7+123 x-3 x^2+17 x^3\big) H_{1,0,1}}{64 (-1+x)^3}
\biggr)
\nonumber \\ &&
+\pi ^2 \biggl[
        -\frac{100078387+125792410 x+60503731 x^2}{71663616 (-1+x)^2}
        -\frac{17}{128} H_{1,1,1,1}
        \nonumber \\ &&
        -\zeta_{3} \biggl(
                \frac{6599+36435 x+8073 x^2+4549 x^3}{27648 (-1+x)^3}
                \nonumber \\ &&
                +\frac{\big(
                        335+147 x+813 x^2-143 x^3\big) H_1}{4608 (1-x)^3}
        \biggr)
        \nonumber \\ &&
        -\frac{\big(
                -43983+479108 x+357953 x^2+115742 x^3\big) H_1}{248832 (-1+x)^2 x}
        \nonumber \\ &&
        +\frac{\big(
                -1028+5603 x+12013 x^2+1157 x^3+867 x^4\big) H_{0,1}}{9216 (-1+x)^3 x}
        \nonumber \\ &&
        +\frac{\big(
                1569-1715 x-12797 x^2+307 x^3\big) H_{1,1}}{13824 (-1+x)^2 x}
        \nonumber \\ &&
        -\frac{\big(
                177+146 x+2244 x^2-1034 x^3+27 x^4\big) H_{0,0,1}}{4608 (-1+x)^3 x}
        \nonumber \\ &&
        +\frac{\big(
                -270-1433 x+5199 x^2+2331 x^3+689 x^4\big) H_{0,1,1}}{4608 (-1+x)^3 x}
        \nonumber \\ &&
        +\frac{\big(
                -192-295 x-1719 x^2-21 x^3+67 x^4\big) H_{1,0,1}}{4608 (-1+x)^3 x}
        \nonumber \\ &&
        +\frac{\big(
                117+1106 x+893 x^2+476 x^3\big) H_{1,1,1}}{2304 (-1+x)^2 x}
        \nonumber \\ &&
        +\frac{\big(
                59+423 x+23 x^2+23 x^3\big) H_{0,0,0,1}}{768 (1-x)^3}
        \nonumber \\ &&
        -\frac{\big(
                9+45 x+7 x^2+2 x^3\big) H_{0,0,1,1}}{64 (-1+x)^3}
        -\frac{\big(
                59+447 x+109 x^2+33 x^3\big) H_{0,1,0,1}}{1536 (-1+x)^3}
        \nonumber \\ &&
        -\frac{\big(
                23+219 x+21 x^2+25 x^3\big) H_{0,1,1,1}}{256 (-1+x)^3}
        -
        \frac{\big(
                1+9 x+x^2+x^3\big) H_{1,0,0,1}}{32 (-1+x)^3}
        \nonumber \\ &&
        -\frac{\big(
                19+183 x+17 x^2+21 x^3\big) H_{1,0,1,1}}{384 (-1+x)^3}
        -\frac{\big(
                23+219 x+21 x^2+25 x^3\big) H_{1,1,0,1}}{768 (-1+x)^3}
\biggr]
\nonumber \\ &&
-\frac{(-8459523+13259174 x) H_1}{2985984 x}
-\frac{\big(
        7125-3574 x-11219 x^2\big) H_{1,1,1}}{3456 (1-x) x}
\nonumber \\ &&
-\frac{\big(
        -422544+762395 x+879722 x^2+1494251 x^3\big) H_{0,1}}{331776 (-1+x)^2 x}
\nonumber \\ &&
-\frac{\big(
        -7560+266112 x+108349 x^2+990011 x^3\big) H_{1,1}}{165888 (-1+x) x^2}
\nonumber \\ &&
-\frac{\big(
        -4440-16709 x-45452 x^2+10171 x^3\big) H_{0,0,1}}{6912 (-1+x)^2 x}
\nonumber \\ &&
-\frac{\big(
        18+930 x-8811 x^2-8782 x^3+2299 x^4\big) H_{0,1,1}}{1152 (-1+x)^2 x^2}
\nonumber \\ &&
-\frac{\big(
        54-5736 x+38353 x^2-10940 x^3+23143 x^4\big) H_{1,0,1}}{6912 (-1+x)^2 x^2}
\nonumber \\ &&
-\frac{\big(
        420+3485 x+13593 x^2-9713 x^3+1521 x^4\big) H_{0,0,0,1}}{2304 (-1+x)^3 x}
\nonumber \\ &&
+\frac{\big(
        366-1722 x-4797 x^2+1084 x^3+416 x^4\big) H_{0,0,1,1}}{576 (-1+x)^3 x}
\nonumber \\ &&
-\frac{\big(
        450-2581 x+1089 x^2-8599 x^3+335 x^4\big) H_{0,1,0,1}}{2304 (-1+x)^3 x}
\nonumber \\ &&
+\frac{\big(
        -1752+1163 x+5102 x^2+3191 x^3\big) H_{0,1,1,1}
}{1152 (-1+x)^2 x}
\nonumber \\ &&
-
\frac{\big(
        -93-934 x-4054 x^2+545 x^3\big) H_{1,0,0,1}}{1152 (-1+x)^2 x}
\nonumber \\ &&
+\frac{\big(
        -101+323 x+42 x^2+216 x^3\big) H_{1,0,1,1}}{192 (-1+x)^2 x}
\nonumber \\ &&
-\frac{\big(
        -186+227 x-700 x^2+677 x^3\big) H_{1,1,0,1}}{1152 (-1+x)^2 x}
+\frac{\big(
        249-98 x+497 x^2\big) H_{1,1,1,1}}{144 (-1+x) x}
\nonumber \\ &&
-\frac{\big(
        12+233 x+1137 x^2-400 x^3+104 x^4\big) H_{0,0,0,0,1}}{384 (-1+x)^3 x}
\nonumber \\ &&
-\frac{\big(
        -60-163 x-423 x^2+158 x^3+26 x^4\big) H_{0,0,0,1,1}}{192 (-1+x)^3 x}
\nonumber \\ &&
-\frac{\big(
        6+6 x+570 x^2-77 x^3+47 x^4\big) H_{0,0,1,0,1}}{192 (-1+x)^3 x}
\nonumber \\ &&
+\frac{\big(
        15+379 x-21 x^2-622 x^3+3 x^4\big) H_{0,0,1,1,1}}{96 (-1+x)^3 x}
\nonumber \\ &&
-\frac{\big(
        -6+79 x+723 x^2+33 x^3+53 x^4\big) H_{0,1,0,0,1}}{192 (-1+x)^3 x}
\nonumber \\ &&
+\frac{\big(
        24+13 x+375 x^2-379 x^3+27 x^4\big) H_{0,1,0,1,1}}{192 (-1+x)^3 x}
\nonumber \\ &&
+\frac{\big(
        -12-517 x+495 x^2-82 x^3+122 x^4\big) H_{0,1,1,0,1}}{384 (-1+x)^3 x}
\nonumber \\ &&
+\frac{\big(
        18-133 x-364 x^2+47 x^3\big) H_{0,1,1,1,1}}{48 (-1+x)^2 x}
\nonumber \\ &&
+\frac{\big(
        9+x+3 x^2+165 x^3+2 x^4\big) H_{1,0,0,0,1}}{192 (-1+x)^3 x}
\nonumber \\ &&
+\frac{\big(
        18-148 x+309 x^2-90 x^3+91 x^4\big) H_{1,0,0,1,1}
}{96 (-1+x)^3 x}
\nonumber \\ &&
+
\frac{\big(
        4+9 x+42 x^2+9 x^3-4 x^4\big) H_{1,0,1,0,1}}{64 (1-x)^3 x}
+\frac{(1+5 x) \big(
        4-10 x+5 x^2\big) H_{1,0,1,1,1}}{16 (1-x)^2 x}
\nonumber \\ &&
-\frac{\big(
        6+29 x+107 x^2-16 x^3\big) H_{1,1,0,0,1}}{192 (1-x)^2 x}
\nonumber \\ &&
+\frac{\big(
        6+160 x-155 x^2+115 x^3\big) H_{1,1,0,1,1}}{96 (1-x)^2 x}
\nonumber \\ &&
+\frac{\big(
        151-14 x+79 x^2\big) H_{1,1,1,0,1}}{96 (-1+x)^2}
+\frac{5 (9+26 x) H_{1,1,1,1,1}}{48 x}
\nonumber \\ &&
-\frac{\big(
        2+24 x+15 x^2+4 x^3\big) H_{0,0,0,0,0,1}}{32 (-1+x)^3}
+\frac{\big(
        7+57 x-3 x^2+5 x^3\big) H_{0,0,0,0,1,1}}{16 (-1+x)^3}
\nonumber \\ &&
-\frac{\big(
        5+36 x+2 x^2+2 x^3\big) H_{0,0,0,1,0,1}}{32 (-1+x)^3}
+\frac{\big(
        23+99 x+29 x^2+5 x^3\big) H_{0,0,0,1,1,1}}{16 (-1+x)^3}
\nonumber \\ &&
-\frac{\big(
        1+12 x+6 x^2+2 x^3\big) H_{0,0,1,0,0,1}}{32 (-1+x)^3}
+\frac{\big(
        4+33 x+2 x^2+3 x^3\big) H_{0,0,1,0,1,1}}{16 (-1+x)^3}
\nonumber \\ &&
-\frac{\big(
        14+48 x+7 x^2\big) H_{0,0,1,1,0,1}}{32 (-1+x)^3}
+\frac{3 (1+x) (1+2 x) H_{0,0,1,1,1,1}}{2 (-1+x)^3}
\nonumber \\ &&
-\frac{x \big(
        3+5 x+x^2\big) H_{0,1,0,0,0,1}}{32 (-1+x)^3}
-\frac{3 \big(
        1+2 x+3 x^2\big) H_{0,1,0,0,1,1}}{16 (-1+x)^3}
\nonumber \\ &&
-
\frac{\big(
        5+27 x+15 x^2+x^3\big) H_{0,1,0,1,0,1}}{64 (-1+x)^3}
-\frac{\big(
        -11+9 x-17 x^2+7 x^3\big) H_{0,1,0,1,1,1}}{32 (-1+x)^3}
\nonumber \\ &&
+\frac{\big(
        2+18 x-x^2+2 x^3\big) H_{0,1,1,0,0,1}}{32 (-1+x)^3}
-\frac{\big(
        2+18 x-x^2+2 x^3\big) H_{0,1,1,0,1,1}}{16 (-1+x)^3}
\nonumber \\ &&
-\frac{3 \big(
        3+15 x+5 x^2+x^3\big) H_{0,1,1,1,0,1}}{32 (-1+x)^3}
+\frac{\big(
        1+9 x+x^2+x^3\big) H_{1,0,0,0,0,1}}{16 (-1+x)^3}
\nonumber \\ &&
+\frac{3 \big(
        1+9 x+x^2+x^3\big) H_{1,0,0,0,1,1}}{8 (-1+x)^3}
+\frac{3 (1+x) (1+2 x) H_{1,0,0,1,1,1}}{4 (-1+x)^3}
\nonumber \\ &&
+\frac{\big(
        3+15 x+5 x^2+x^3\big) H_{1,0,1,0,1,1}}{16 (-1+x)^3}
-\frac{\big(
        2+12 x+3 x^2+x^3\big) H_{1,0,1,1,0,1}}{8 (-1+x)^3}
\nonumber \\ &&
-\frac{3}{4} H_{1,0,1,1,1,1}
+\frac{\big(
        1+9 x+x^2+x^3\big) H_{1,1,0,0,0,1}}{16 (-1+x)^3}
+\frac{(1+x) (1+2 x) H_{1,1,0,0,1,1}}{4 (-1+x)^3}
\nonumber \\ &&
-\frac{\big(
        3+15 x+5 x^2+x^3\big) H_{1,1,0,1,0,1}}{32 (-1+x)^3}
-\frac{9}{16} H_{1,1,0,1,1,1}
-\frac{3}{8} H_{1,1,1,0,1,1}
\nonumber \\ &&
-\frac{3}{16} H_{1,1,1,1,0,1}
-\frac{15}{8} H_{1,1,1,1,1,1}
-\frac{15}{16} H_{0,1,1,1,1,1}
\nonumber \\ &&
-\frac{\big(
        10+1185 x-42 x^2+197 x^3\big) \zeta_{3}^2}{576 (-1+x)^3}
        ~,
\end{eqnarray}
where
$\zeta_i$ denote the Riemann $\zeta$ function at integer
argument $i$.
Furthermore, we use the following convention for the
iterated integrals:
\begin{eqnarray}
    H_{i,\vec{w}}(x) &=&
    \int\limits_{0}^{x} {\rm d}t \, w_i(t) \, H_{\vec{w}}(t)~,
\end{eqnarray}
with the letters
\begin{align*}
    w_{0}(t) &= \frac{1}{t} ~, &
    w_{-1}(t) &= \frac{1}{1+t} ~, &
    w_{1}(t) &= \frac{1}{1-t} ~, &
    w_{2}(t) &= \frac{1}{2-t} ~,
\end{align*}
and we drop the argument for brevity, i.e.
$H_{\vec{w}}(x) \equiv H_{\vec{w}}$.
The first three letters define the harmonic polylogarithms.
The forth letter can be avoided by allowing for
harmonic polylogarithms evaluated at argument $1-x$.

We compared our analytic results for $F_1^v$, $F_2^v$, $F_3^v$ and $F^s$
to the ones attached to Ref.~\cite{Datta:2023otd}
including $\epsilon^4$
and $\epsilon^2$ terms at one and two-loop order, respectively.
We found full agreement
after adjusting for the different tensor basis and
renormaliztion and after adapting the large-$N_C$ limit and setting
all fermionic contributions to zero.


\section{Numerical results}
\label{sec::results_num}

As mentioned in Section~\ref{sub::MIs} we compute all master integrals
using the method ``expand and match''. As a result we obtain
analytic expansions of the (unknown) three-loop expressions around the
values $s/m^2$ given in Eq.~(\ref{eq::som2}) with high-precision numerical
coefficients. Note that our approach provides generalized expansions
which may contain logarithms of square roots of the expansion
parameter, depending on the physical situation at the expansion point.

To illustrate the structure of our results we show in the following the
first three expansion terms for $s/m^2\to 1$
of the $C_F^3$ colour factor of the renormalized and infrared subtracted form factor $C_1^t$.
It is given by\footnote{We truncate the numerical values to six significant digits and suppress trailing zeros.}
\begin{align}
C_1^{t,(3)}\Big|_{C_F^3} &=
    3.95625
    \notag \\ &
    + 1.23578 \lb - 1.02622 \lb^2 + 1.06563 \lb^3 - 0.37851 \lb^4 + 0.0625 \lb^5 - 0.0208333 \lb^6
    \notag \\&
    + \b (8.31567 + 3.58274 \lb - 2.03938 \lb^2 - 0.0683922 \lb^3 - 0.4375 \lb^4 + 0.125 \lb^5)
    \notag \\&
    + \b^2 (-3.51595 - 19.1367 \lb + 4.25689 \lb^2 + 1.3063 \lb^3 + 0.614583 \lb^4 - 0.1875 \lb^5)
    \notag \\&
    + \b^3 (9.9209 + 35.8225 \lb - 2.05302 \lb^2 - 4.15664 \lb^3 + 0.194444 \lb^4 + 0.125 \lb^5)
    \notag \\&
  +O(\beta^4),
  \label{eq:Ct1-s1}
\end{align}
where $\beta=(1-m^2/s)/(1+m^2/s)$ and $L_\beta = \log(-2 \beta)$.
One observes that the expansion is logarithmically divergent in the
limit $\beta \to 0$, however, it does not contain power suppressed terms
like $1/\beta^n$, which are present in the bare amplitude.
Similarly, we have a power-log expansion around $s/m^2 = -\infty$.
The expansion around the other $s/m^2$ values are all
simple Taylor expansions.

We implement the expansions around the $s/m^2$ values of Eq.~(\ref{eq::som2})
in a {\tt Fortran} program {\tt FFh2l} which can be obtained from the
website~\cite{FFh}.  It is either possible to access the three-loop
expressions within {\tt Fortran} or via a {\tt Mathematica} interface which
has the same functionality. {\tt FFh2l} provides results for the pole parts
and finite contributions of all twelve ultraviolet renormalized form factors
but also for the finite parts of the infrared subtracted form factors $C$.
In the region
\begin{equation}
  -75 < s/m^2 < 15/16
  \label{eq:grid-region}
\end{equation}
we provide a grid by numerically evaluating our Taylor expansions and the analytic counterterms with the help of \texttt{GiNaC}~\cite{Bauer:2000cp,Vollinga:2004sn}.
Around the singular points $s/m^2 \to -\infty$ and $s/m^2 = 1$ we switch to dedicated power-log expansions as shown in Eq.~(\ref{eq:Ct1-s1}).
This includes expansions of the counterterms to increase stability.
A more detailed description of {\tt FFh2l} can be found in Appendix~\ref{app::code}.

As reference, we show in Fig.~\ref{fig::FF} the (finite) vector,
scalar and tensor form factors for $\mu^2=m^2$ as a function of
$s/m^2$ for $0<s<m^2$.
\begin{figure}[t]
  \begin{center}
    \includegraphics[width=0.45\textwidth]{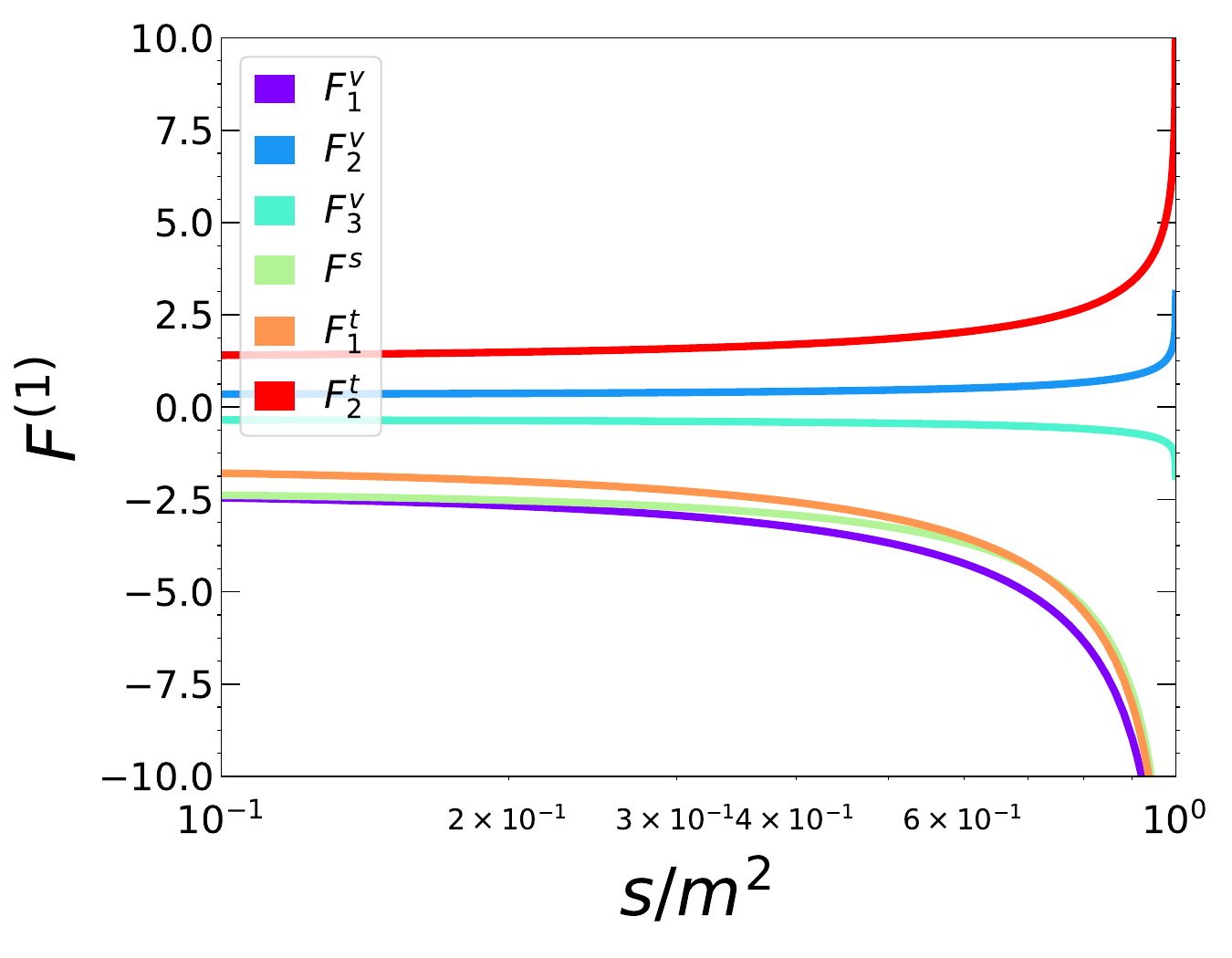}
    \includegraphics[width=0.45\textwidth]{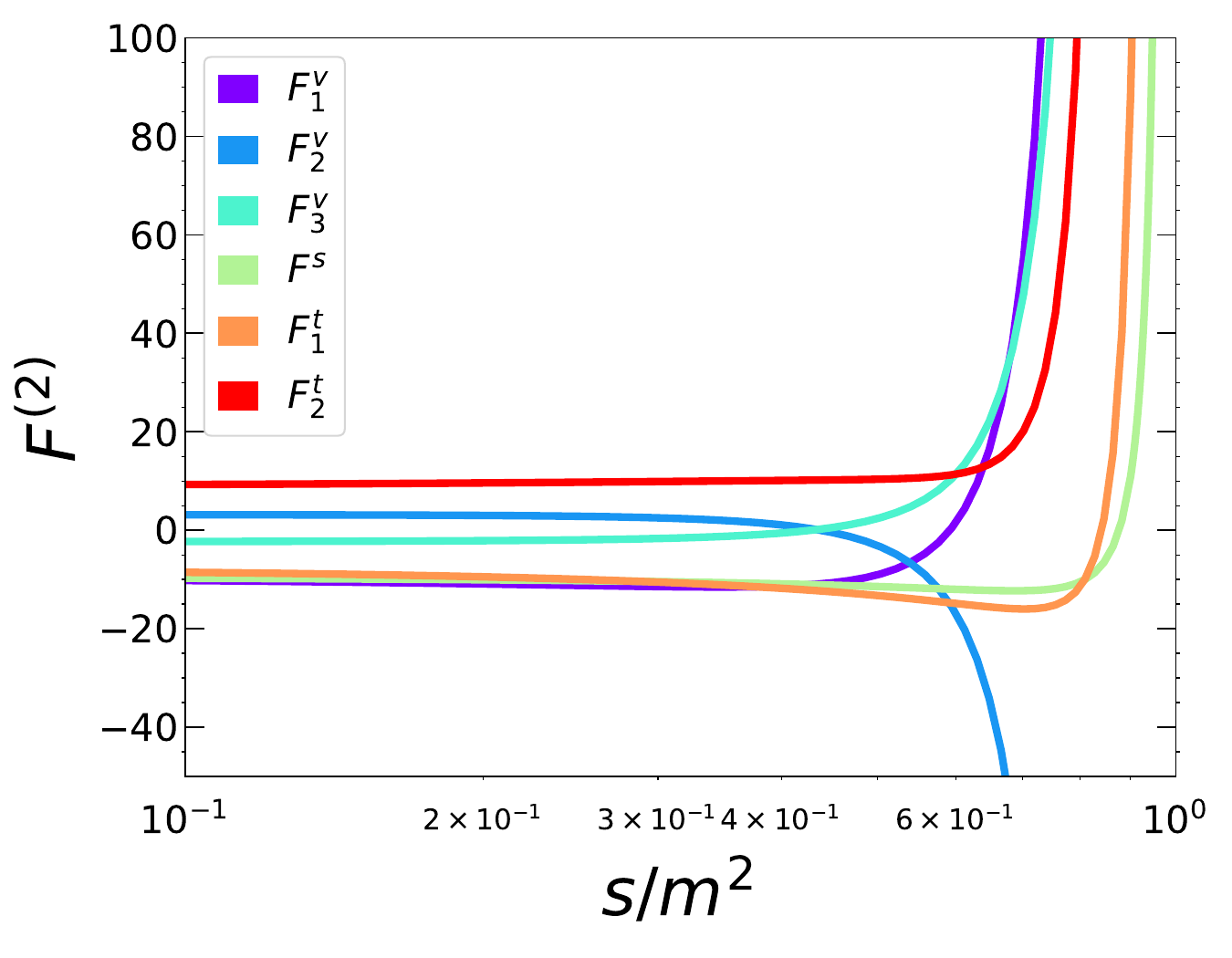}
    \\
    \includegraphics[width=0.45\textwidth]{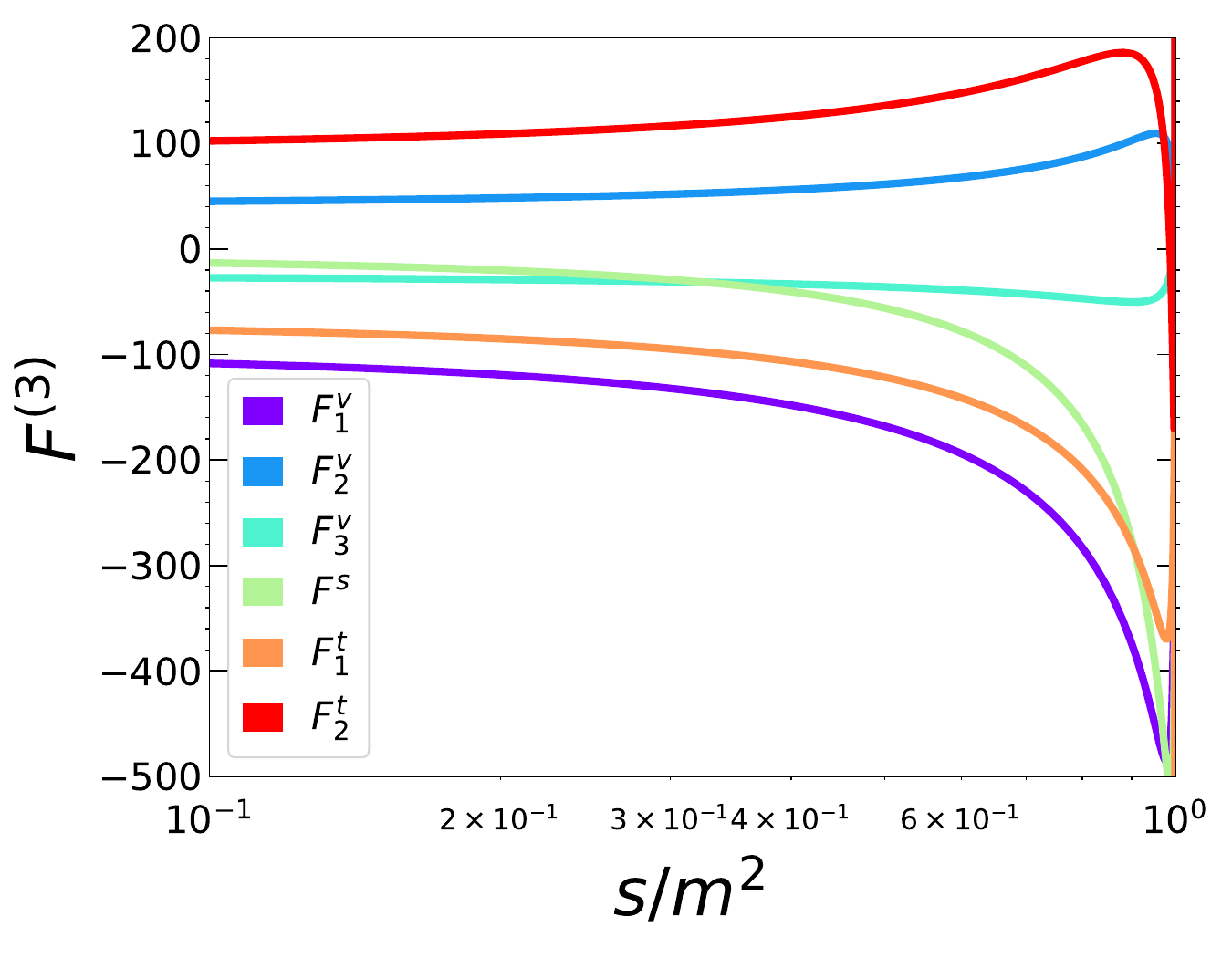}
  \caption{\label{fig::FF}One-, two-, and three-loop form factors as a function of $s/m^2$ for $s>0$. The colour factors have been adjusted to QCD with $n_l=4$ and $n_h=1$. For the renormalization scale $\mu^2=m^2$ has been chosen.}
  \end{center}
\end{figure}
We remind the reader that the axialvector and pseudoscalar form factors are related to the vector form factors through Eq.~(\ref{eq:chiral-symmetry}) and that $C^{t}_3 = C^{t}_4 = 0$ as discussed in Section~\ref{sec::IR}.
For the colour factor we have chosen $C_A=3$, $C_F=4/3$ and
$T_F=1/2$. Furthermore, we have $n_l=4$ and $n_h=1$.  For the $x$ axis
we have chosen a logarithmic scale since there is only a mild
variation of the from factors for $s\approx 0$.  On the other hand,
at all loop orders we observe Coulomb-like singularities close to
threshold.  It is straightforward to reproduce these plots by either
using the analytic one- and two-loop expressions provided as an
ancillary file or with the help of the package {\tt FFh2l}.

There are several checks on the correctness of our calculation.  First of all,
we observe that the gauge parameter cancels in the ultraviolet renormalized
expressions. The analytic contributions induced by the one- and two-loop
results cancel against the numerical results from the bare three-loop
form factors. We observe that this cancellation happens at the level
of $10^{-23}$ or significantly better which at the same time
is an indication for the precision of our semi-analytic three-loop result.

\begin{figure}[t]
  \begin{center}
    \begin{tabular}{cc}
      \includegraphics[width=0.47\textwidth]{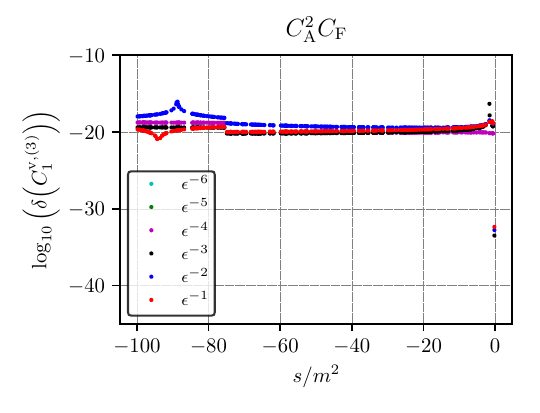}
      &
      \includegraphics[width=0.47\textwidth]{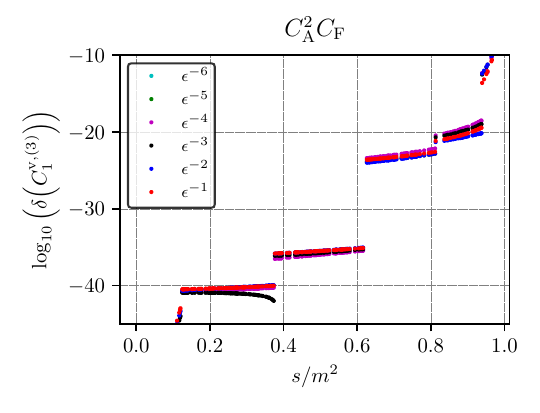}
      \\
      \includegraphics[width=0.47\textwidth]{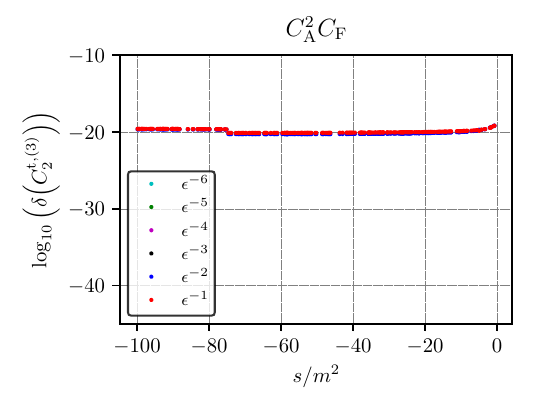}
      &
      \includegraphics[width=0.47\textwidth]{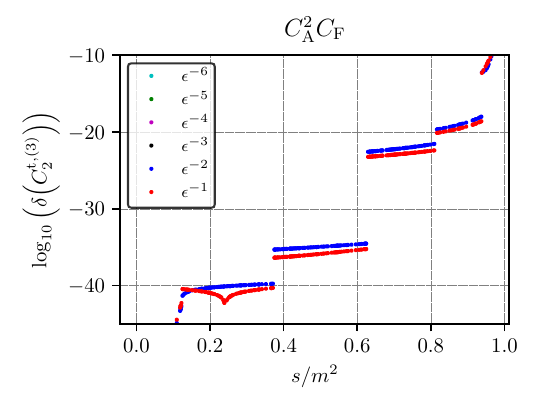}
      \\
    \end{tabular}
    \caption{
      \label{fig:pole-cancellation}
      Relative cancellation of the poles for the the $C_A^2C_F$ color structures of $C_1^{v,(3)}$ and $C_2^{t,(3)}$. They show the worst
      behaviour of all form factors. The left panels cover the region $[-\infty,0]$ and on the right results for $0 < s < 1$ are shown.}
  \end{center}
\end{figure}

An important check is the cancellation of the $1/\epsilon$ poles
in the construction of $C$.
As
expected, there are poles up to $1/\epsilon^6$. All of them cancel after
ultraviolet renormalization and infrared subtraction.
Here, we proceed as in Refs.~\cite{Fael:2022miw,Fael:2023zqr}
and define
\begin{equation}
  \label{eq::delta-def}
  \delta \Big(C^{(3)}\big|_{\ep^i}\Big) = \frac{F^{(3)}\big|_{\ep^i}+F^{(\text{CT}+{Z})}\big|_{\ep^i}}{F^{(\text{CT}+{Z})}\big|_{\ep^i}}\,,
\end{equation}
where $F^{(3)}$ stands for the bare three-loop contribution and
$F^{(\text{CT}+{Z})}$ contains the contributions induced from the analytic
tree-level, one- and two-loop terms due to ultraviolet renormalization
(``CT'') and infrared subtraction (``Z'').  In the region given by Eq.~(\ref{eq:grid-region}), we observe that there is a
cancellation of at least 16 digits for each individual colour of each form factor and each $\epsilon$ pole.
Only for $s/m^2 > 15/16$ the cancellation of the grid drops below that due to the Coulomb-like singularity which supports our decision to switch to a dedicated expansion.
In most parts of the phase space the cancellation is many orders of magnitude better as can be seen in Fig.~\ref{fig:pole-cancellation} where we show the two worst cases of all form factors.

Remarkably, all six orders of $\epsilon$ cancel with a similar precision.
Only a careful analysis reveals a slight trend towards worse precision for the lower poles.
Especially in the region $0 < s/m^2 < 1$ the loss of precision when switching to the next expansion point is clearly visible, but remains on a very high level.
On the negative axis, the precision curve is much smoother and only the matching from our boundary conditions at $s/m^2 = 0$ to $s/m^2 = -1/2$ and the matching from $s/m^2 = -60$ to $s/m^2 \to -\infty$ stick out.

Finally, we can also check the Ward identity from Eq.~(\ref{eq:Ward-identity}).
Naively, one would expect that it allows us to estimate the precision of the finite terms similar to the pole cancellation.
However, this is not the case.
It was noticed in the two-loop calculation of Ref.~\cite{Bonciani:2008wf} that the Ward identity is fulfilled already on the level of the master integrals.
We observe something similar: the sum of the bare three-loop contributions and the sum of the counterterm contributions to Eq.~(\ref{eq:Ward-identity}) are separately constant, but nonzero, and vanish when summing both contributions.
This suggests that there is a similar relation between the master integrals also at the three-loop level.
Since in our calculation we do not express the renormalization constants in terms of master integrals,
we check Eq.~(\ref{eq:Ward-identity}) only numerically and
observe that it is fulfilled to high
precision.
In most parts of the phase space it exceeds our internal precision of $50$ digits and only rarely drops below that at less stable points.
Even at $s/m^2 \approx 0.9374$, where we switch to the dedicated power-log expansion due to the Coulomb-like singularity at the threshold, the Ward identity holds to at least $19$ digits.

After all these considerations, we estimate the precision of the finite terms by extrapolating the pole cancellations and expect that our result is correct to at least $14$ digits in the grid region given by Eq.~(\ref{eq:grid-region}) and usually many more in most parts of the phase space.

For the two singular power-log expansions around $s/m^2 \to -\infty$ and $s/m^2 = 1$ our strategy to estimate their precision differs slightly.
As mentioned before, here we also expand the counterterms to increase stability.
Hence, we can check the cancellation of the $1/\epsilon$ poles order by order in the expansion parameters $-m^2/s$ and $(1-s/m^2)$, respectively.
For the expansion around $s/m^2 \to -\infty$, we observe that they cancel with at least $15$ digits up to order $(-m^2/s)^5$ and with at least $10$ digits up to $(-m^2/s)^{17}$.
The expansion around $s/m^2 = 1$ behaves worse and the coefficients cancel with at least $17$ digits up to order $(1-s/m^2)^1$, with at least $10$ digits up to $(1-s/m^2)^3$, and with at least $9$ digits up to $(1-s/m^2)^{20}$.
Similarly, we can also check the Ward identity~(\ref{eq:Ward-identity}) order by order in the expansion parameters.
Again we observe that it holds with high precision, reaching our internal precision of $50$ digits for most expansion orders.
Hence, we conservatively estimate that the two power-log expansions in the singular regions are sufficient to provide $10$ correct digits for the finite part.

With this in mind, the grids and expansions provided in {\tt FFh2l} are designed to provide at least $10$ correct digits over the full range $-\infty < s/m^2 < 1$.


\section{The hard function in \bsgamma}
\label{sec:hardfunctionbsgamma}

In a SCET-based approach to \bsgamma~the decay width is written as the product of a hard function with a convolution of the jet and soft function~\cite{Korchemsky:1994jb,Akhoury:1995fp,Neubert:2004dd}. While the latter two are known to three loops already~\cite{Becher:2005pd,Becher:2006qw,Bruser:2018rad,Bruser:2019yjk} the hard function was up to now only known to two loops~\cite{Ali:2007sj,Ligeti:2008ac,Dehnadi:2022prz}. With the three-loop matching coefficients of the tensor current at hand, we are now in the position to extract the hard function of \bsgamma~to three loops as well.

To this end, we follow the discussions in Refs.~\cite{Ali:2007sj,Ligeti:2008ac,Dehnadi:2022prz} and consider the operator
\begin{equation}
Q_7 = -\frac{e \, \overline m_b(\mu)}{4 \pi^2} \, (\bar s_L \sigma_{\mu\nu} F^{\mu\nu} b_R) \, , \label{eq:Q7}
\end{equation}
where $\overline m_b(\mu)$ is the bottom-quark mass in the $\overline{\rm{MS}}$ scheme and $e$ the electric charge of the positron. At leading power this operator is matched onto the SCET current
\begin{equation}
J^A  = (\bar \xi W_{hc}) /\!\!\!\epsilon_\perp (1-\gamma_5) h_v \, ,  \label{eq:JA}
\end{equation}
with the heavy quark effective theory (HQET) field $h_v$ of the heavy quark, the SCET field $\xi$ of the light quark, the hard-collinear Wilson line $W_{hc}$ and the polarization vector $\epsilon_\perp^\mu$ of the on-shell photon. The field strength tensor $F^{\mu\nu}$ in Eq.~\eqref{eq:Q7} gives rise to the Feynman rule
\begin{equation}
F^{\mu\nu} = \partial^\mu A^\nu - \partial^\nu A^\mu \longrightarrow i \left( q^\mu \epsilon_\perp^\nu - q^\nu \epsilon_\perp^\mu \right) \, . \label{eq:Feynmanrule}
\end{equation}
If the matching is done on-shell, one can use $\epsilon_\perp \cdot q_2 = \epsilon_\perp \cdot q_1 = 0$, and arrive for $q^2=0$ at
\begin{equation}
\langle s \gamma | Q_7 | b \rangle = -\frac{e \, \overline m_b \, 2 E_\gamma}
{4 \pi^2} \, \left(F_1^t - \frac{1}{2} \, F_2^t - \frac{1}{2} \, F_3^t\right)_{\big|q^2=0} \, \times J^A \, ,
\end{equation}
where $2E_\gamma \approx m_b$ at leading power. After infrared subtraction the expression in parenthesis becomes
\begin{equation}
C_\gamma \, \equiv \, C_1^{t}(s=0) - \frac{1}{2} \, C_2^{t}(s=0) \, .
\label{eq:matchingbsg}
\end{equation}
The factorization formula of \bsgamma~is formulated on the level of the decay rate. Moreover, since the hard function $h_s(\mu)$ in \bsgamma~is a genuine SCET object, the logarithms of the QCD scale $\nu$ have to be set to zero in the following. We therefore arrive at
\begin{align}
 h_s(\mu) & = \Big| {C_{\gamma}}_{\big| L_\nu=0}\Big|^2 \, .
\end{align}

The explicit result of $h_s(\mu)$ to three loops reads
\begin{align}
h_s(\mu) & = 1 +  C_F \left[-L_\mu^2-5
   L_\mu-\frac{\pi^2}{6}-12\right] \left(\frac{\alpha_s^{(n_l)}(\mu)}{4\pi}\right) \nonumber \\[0.6em]
   &+ \left[\frac{1}{2}C_F^2 L_\mu^4 + L_\mu^3 \left(-\frac{11}{9}C_A C_F+5 C_F^2+\frac{4}{9} C_F n_l
   T_F\right) \right. \nonumber \\[0.6em]
   &  \hspace*{15pt}+ \left(\left(\frac{\pi ^2}{3}-\frac{299}{18}\right) C_A C_F+\left(\frac{49}{2}+\frac{\pi ^2}{6}\right) C_F^2+\frac{50}{9}  C_F n_l T_F \right) L_\mu^2 \nonumber \\[0.6em]
   & \hspace*{15pt} + \left(C_A C_F \left(22 \zeta_3-\frac{3925}{54}-\frac{16 \pi ^2}{9}\right)+C_F^2 \left(-24 \zeta_3 +\frac{117}{2}+\frac{17 \pi ^2}{6}\right) \right. \nonumber \\[0.6em]
   & \left.\hspace*{40pt} +\left(\frac{682}{27}+\frac{8 \pi ^2}{9}\right) C_F n_l T_F\right) L_\mu +C_F T_F \left(\frac{7126}{81}-\frac{16 \zeta_3}{3}-\frac{232 \pi ^2}{27}\right)\nonumber \\[0.6em]
   &  \hspace*{15pt} +C_A C_F \left(-\frac{122443}{648}+\frac{478 \zeta_3}{9}+\frac{829 \pi ^2}{108}+\frac{31 \pi^4}{60}-\frac{74}{3} \pi ^2 \ln(2)\right) \nonumber \\[0.6em]
   &\hspace*{15pt} +C_F^2 \left(\frac{3379}{24}-88 \zeta_3-25 \pi ^2-\frac{47 \pi ^4}{72}+\frac{148}{3} \pi ^2 \ln(2)\right) \nonumber \\[0.6em]
   &  \left. \hspace*{15pt} +C_F n_l T_F \left(\frac{52 \zeta_3}{9}+\frac{7859}{162}+\frac{109 \pi ^2}{27}\right)\right] \left(\frac{\alpha_s^{(n_l)}(\mu)}{4\pi}\right)^{\!\!2} \nonumber \\[0.6em]
   & +\left[-\frac{1}{6}C_F^3 L_\mu^6 +\left(-\frac{5}{2}  C_F^3+\frac{11}{9} C_A C_F^2 -\frac{4}{9} n_l T_F C_F^2\right) L_\mu^5 +\left(-\frac{121}{54}  C_A^2 C_F-\frac{70}{9} n_l T_F C_F^2 \right. \right. \nonumber \\[0.6em]
   & \left.  \hspace*{40pt} -\left(\frac{37}{2}+\frac{\pi^2}{12}\right) C_F^3+\left(\frac{409}{18}-\frac{\pi^2}{3}\right) C_A C_F^2-\frac{8}{27} n_l^2 T_F^2 C_F+\frac{44}{27} C_A n_l T_F C_F\right) L_\mu^4 \nonumber \\[0.6em]
   & \hspace*{15pt} +\left(\left(24 \zeta_3-\frac{238}{3}-\frac{17 \pi^2}{6}\right) C_F^3- \left(\frac{1540}{27}+\frac{26 \pi^2}{27}\right) n_l T_F C_F^2 -\frac{400}{81} n_l^2 T_F^2 C_F\right. \nonumber \\[0.6em]
   & \hspace*{40pt} + \left(\frac{4601}{27}+\frac{17 \pi^2}{54}-22 \zeta_3\right) C_A C_F^2+ \left(\frac{2476}{81} -\frac{8 \pi^2}{27}\right) C_A n_l T_F C_F \nonumber \\[0.6em]
   & \hspace*{40pt} \left. +\left(\frac{22 \pi^2}{27} -\frac{3595}{81}\right) C_A^2 C_F\right) L_\mu^3  \nonumber \\[0.6em]
   & \hspace*{15pt} +\left( \left(-\frac{6799}{24}+\frac{155 \pi^2}{12}+\frac{47 \pi^4}{72}-\frac{148}{3} \pi^2 \ln(2)+208 \zeta_3\right) C_F^3 \right. \nonumber \\[0.6em]
   & \hspace*{40pt} +\left(\frac{92 \zeta_3}{9}-\frac{34205}{162}-\frac{326\pi^2}{27}\right) n_l T_F C_F^2 + \left(\frac{16 \zeta_3}{3}-\frac{7126}{81}+\frac{232 \pi^2}{27}\right) T_F C_F^2 \nonumber \\[0.6em]
   & \hspace*{40pt} + \left(\frac{483547}{648}+\frac{395 \pi^2}{54}-\frac{103 \pi^4}{180}+\frac{74}{3} \pi^2
   \ln(2)-\frac{2260 \zeta_3}{9}\right) C_A C_F^2 \nonumber \\[0.6em]
   & \hspace*{40pt} - \left(\frac{2680}{81}+\frac{32 \pi^2}{27}\right) n_l^2 T_F^2 C_F  + \left(\frac{220 \zeta_3}{3}-\frac{27190}{81}-\frac{14 \pi^2}{9}-\frac{11 \pi^4}{45}\right) C_A^2 C_F \nonumber \\[0.6em]
   & \hspace*{40pt} \left. + \left(\frac{17956}{81}+\frac{112 \pi^2}{27}-\frac{32 \zeta_3}{3}\right) C_A n_l T_F C_F\right) L_\mu^2 \nonumber \\[0.6em]
   & \hspace*{15pt} +\left(\left(-\frac{16811}{24}+\frac{393 \pi^2}{4}+\frac{479 \pi^4}{360}-\frac{740}{3} \pi^2
   \ln(2)+660 \zeta_3+\frac{28 \pi^2 \zeta_3}{3}+240 \zeta_5\right) C_F^3  \right. \nonumber \\[0.6em]
   & \hspace*{40pt} +
   \left(\frac{1479851}{648}-\frac{29185 \pi^2}{162}-\frac{2887 \pi^4}{540}+\frac{4366}{9} \pi^2 \ln(2)-\frac{13106 \zeta_3}{9} -120 \zeta_5 \right. \nonumber \\[0.6em]
   & \hspace*{65pt}\left. -\frac{19 \pi^2 \zeta_3}{3} \right) C_A C_F^2 +
   \left(\frac{80 \zeta_3}{3}-\frac{35630}{81}+\frac{1160 \pi^2}{27}\right) T_F C_F^2\nonumber \\[0.6em]
   & \hspace*{40pt}  + \left(\frac{692 \zeta_3}{3}-\frac{86683}{162}+\frac{2812\pi^2}{81}+\frac{16 \pi^4}{27}-\frac{1184}{9} \pi^2 \ln(2)\right) n_l T_F C_F^2  \nonumber \\[0.6em]
   & \hspace*{40pt} + \left(-\frac{1171918}{729}+\frac{12374 \pi^2}{243}+\frac{107 \pi^4}{45}-\frac{1628}{9} \pi^2 \ln(2)+\frac{18874 \zeta_3}{27}-100 \zeta_5 \right. \nonumber \\[0.6em]
   & \hspace*{65pt} \left.-\frac{56 \pi^2 \zeta_3}{9}\right) C_A^2 C_F - \left(\frac{83776}{729}+\frac{992 \pi^2}{81}+\frac{448 \zeta_3}{27}\right) n_l^2 T_F^2 C_F \nonumber \\[0.6em]
   & \hspace*{40pt}  + \left(\frac{156772}{243}-\frac{5104 \pi^2}{81}-\frac{352\zeta_3}{9}\right) C_A T_F C_F  \nonumber \\[0.6em]
   & \hspace*{40pt}  + \left(\frac{128 \zeta_3}{9}-\frac{57008}{243}+\frac{1856\pi^2}{81}\right) n_l T_F^2 C_F \nonumber \\[0.6em]
   & \hspace*{40pt} \left . + \left(\frac{677290}{729}+\frac{4364 \pi^2}{243}-\frac{8 \pi^4}{9}+\frac{592}{9} \pi^2 \ln(2)-\frac{1040 \zeta_3}{9}\right) C_A n_l T_F C_F\right) L_\mu  \nonumber \\[0.6em]
   & \hspace*{15pt} + \left(\frac{175459 \pi^2}{972}-\frac{219365}{486}-\frac{41303 \pi^4}{2430}-\frac{3776}{9} \pi^2 \ln(2)+\frac{1472}{27} \pi^2 \ln^2(2) \right. \nonumber \\[0.6em]
   & \hspace*{40pt} \left. +\frac{1072 \ln^4(2)}{27}+\frac{8576 \mathrm{Li}_4\!\left(\frac{1}{2}\right)}{9}+\frac{64816 \zeta_3}{81}+\frac{298 \pi^2 \zeta_3}{9}+\frac{896 \zeta_5}{9}\right) C_F^2 n_l T_F \nonumber \\[0.6em]
   & \hspace*{15pt} + \left(\frac{8584738}{6561}+\frac{151303 \pi^2}{2187}+\frac{4703 \pi^4}{1215}+\frac{1888}{9} \pi^2 \ln(2)-\frac{736}{27} \pi^2 \ln^2(2)-\frac{536 \ln^4(2)}{27} \right.\nonumber \\[0.6em]
   & \hspace*{40pt} \left. -\frac{4288\mathrm{Li}_4\!\left(\frac{1}{2}\right)}{9}-\frac{12640 \zeta_3}{81}-\frac{76 \pi^2 \zeta_3}{9}-136 \zeta_5\right) C_A C_F n_l T_F \nonumber \\[0.6em]
   & \hspace*{15pt}  -95.12984922305611775005 \, C_A C_F T_F  +1429.62034756690622959783 \, C_A C_F^2 \nonumber \\[0.6em]
   & \hspace*{15pt} -3126.14625382895615802902 \, C_A^2 C_F  +181.97737877492915588766 \, C_F^2 T_F \nonumber \\[0.6em]
   & \hspace*{15pt} +345.53350842018910941336 \, C_F^3  +\left(\frac{128 \pi^2}{15}-\frac{23936}{81}-\frac{32 \pi^4}{135}+\frac{1664 \zeta_3}{9}\right) C_F T_F^2  \nonumber \\[0.6em]
   & \hspace*{15pt} +\left(\frac{7088 \pi^2}{243}-\frac{211888}{729}-\frac{64 \pi^4}{405}+\frac{256 \zeta_3}{27}\right) C_F n_l T_F^2  \nonumber \\[0.6em]
   & \hspace*{15pt} \left. -\left(\frac{741898}{6561}+\frac{6632 \pi^2}{243}+\frac{884 \pi^4}{1215}+\frac{20672 \zeta_3}{243}\right) C_F n_l^2 T_F^2 \right] \left(\frac{\alpha_s^{(n_l)}(\mu)}{4\pi}\right)^{\!\! 3} + \mathcal{O}(\alpha_s^4) \, . \label{eq:bsgammahard3loops}
\end{align}
In this expression, the bottom-quark mass in $L_\mu = \ln(\mu^2/m_b^2)$ is renormalized in the pole scheme. In this scheme, the hard function satisfies the following RGE,
\begin{equation}\label{eq:RGEbsghard}
\frac{{\rm d}h_s(\mu)}{{\rm d}\ln\mu}
= \biggl[-\gamma^{\rm cusp}(\alpha_s^{(n_l)}(\mu))\ln\frac{\mu^2}{m_b^2} + 2\gamma^H(\alpha_s^{(n_l)}(\mu))\biggr] h_s(\mu) \, .
\end{equation}
At a given order in $\alpha_s$, all terms containing $L_\mu$ are determined by the anomalous dimension coefficients and lower-loop results, and all our $L_\mu$ terms agree with the derivation in Ref.~\cite{Dehnadi:2022prz}. The $L_\mu$-independent terms at three loops are, however, genuinely new. In Eq.~\eqref{eq:bsgammahard3loops}, all terms through to two loops are analytic and agree with Refs.~\cite{Ali:2007sj,Ligeti:2008ac,Bell:2010mg,Dehnadi:2022prz}. At three loops, all terms containing $L_\mu$, as well as the light fermionic pieces and the color factor $C_F T_F^2$ are also analytic. The remaining ones are obtained numerically to at least 100 decimal digits, of which we display 20 in the present write-up. An electronic version of Eq.~\eqref{eq:bsgammahard3loops} can be downloaded from the webpage~\cite{progdata}.

Upon substituting the numerical values $C_A=3$, $C_F=4/3$, $T_F=1/2$, and $n_l=4$ for the color and flavor factors, the expansion of $h_s$ for $\mu=m_b$ reads
\begin{align}
    h_s(m_b) & = 1
    - 4.5483113556160754788 \left(\!\frac{\alpha_s^{(4)}(m_b)}{\pi}\!\right) -19.286105172591724459 \left(\!\frac{\alpha_s^{(4)}(m_b)}{\pi}\!\right)^{\!\!2}  \nonumber \\[0.6em]
    & \hspace*{22pt}-181.16173810663548219 \left(\!\frac{\alpha_s^{(4)}(m_b)}{\pi}\!\right)^{\!\!3} + \mathcal{O}(\alpha_s^4) \, .
\end{align}
An interesting detail to note is that the coefficient $h_3$, which was treated as a nuisance parameter in Ref.~\cite{Dehnadi:2022prz} and varied in the range $h_3=0\pm 80$, comes out of the genuine three-loop calculation as $h_3 = -181.1617381$ and therefore more than a factor of two larger in magnitude compared to the variation boundaries.


\section{Conclusion}
\label{sec::conclusion}

We compute the three-loop QCD corrections to heavy-to-light transitions for
the entire set of Dirac bilinears which are independent in four space-time
dimensions. The calculations use state-of-the art multi-loop techniques and a
well-established workflow, starting from the generation of the amplitude and
the projection onto Lorentz-covariant structures. The resulting scalar
integrals are subsequently reduced to master integrals. A certain subset of
master integrals (one- and two-loop integrals, three-loop leading color and
fermionic integrals apart from the ones with a single closed heavy fermion
loop) are obtained analytically, while for the others the differential
equations are solved via the ``expand and match'' method, which uses
expansions about several kinematic points and as such gives semi-analytic
results for the form factors.

Infrared subtraction is applied to the ultraviolet-renormalized QCD form factors at three loops, and finite matching coefficients to SCET are obtained. In this procedure, the poles in the dimensional regulator $\epsilon$ cancel to at least $12$ digits and we thus estimate the precision of the finite part to be at least $10$ digits. From the matching coefficients of the tensor current at light-like momentum transfer, the three-loop correction to the hard function in $\bar B \to X_s \gamma$ is extracted. Further phenomenological applications to rare semileptonic decays, top-quark or muon decays are left for future investigations.

Electronic results are provided as {\tt Mathematica} and {\tt Fortran} codes which allow for fast and precise numerical evaluations for physically relevant values of the square of the four-momentum transfer (we do not consider values of $s/m^2 > 1$, though).
The supplemenatary material to this paper can be found on the websites~\cite{progdata,FFh,fael_2024_11046426}.



\section*{Acknowledgements}

We thank Johann Usovitsch and Zihao Wu for allowing us to use the development version of \texttt{Kira} and Ze Long Liu for discussion about the infrared singularity structure. Moreover, we thank Robin Br\"user and Maximilian Stahlhofen for collaboration at initial stages and useful correspondence. The research of T.H., J.M., and M.S.\ was supported by the Deutsche Forschungsgemeinschaft (DFG, German Research Foundation) under grant 396021762 --- TRR 257 ``Particle Physics Phenomenology after the Higgs Discovery''. K.S.\ has
received funding from the European Research Council (ERC) under the
European Union’s Horizon 2020 research and innovation programme grant
agreement 101019620 (ERC Advanced Grant TOPUP). The work of M.F.\ was
supported by the European Union’s Horizon 2020 research and innovation
program under the Marie Sk\l{}odowska-Curie grant agreement
No.~101065445 - PHOBIDE.
The work of F.L.~was supported by the Swiss National Science Foundation (SNSF)
under contract \href{https://data.snf.ch/grants/grant/211209}{TMSGI2\_211209}.
The Feynman diagrams were drawn with the help of Axodraw~\cite{Vermaseren:1994je} and JaxoDraw~\cite{Binosi:2003yf}.


\begin{appendix}


\section{\label{app::proj}Projectors}

The scalar form factors introduced in Eq.~(\ref{eq::Gamma})
are obtained by the application of the appropriate projectors
via
\begin{eqnarray}
  F^\delta_i &=& \mbox{Tr}\left[ P^\mu_{\delta,i} \Gamma^\delta \right]\,,
\end{eqnarray}
where the $P^\mu_{\delta,i}$ are given by
\begin{eqnarray}
  P^\mu_{v,i} &=&
  \slashed{q}_1\left[
    g^v_{1,i} \gamma^\mu +
    g^v_{2,i} \frac{p^\mu}{m} +
    g^v_{3,i} \frac{q^\mu}{m}
    \right] \left(\slashed{q}_2+m\right)
  \,, \nonumber\\
  P^\mu_{a,i} &=&
  \slashed{q}_1\left[
    g^a_{1,i} \gamma^\mu +
    g^a_{2,i}
    {\frac{p^\mu}{m}} +
    g^a_{3,i}
    {\frac{q^\mu}{m}}
    \right] \gamma_5 \left(\slashed{q}_2+m\right)
  \,, \nonumber\\
  P^{s} &=&
\slashed{q}_1 g_s \left(\slashed{q}_2+m\right)
  \,, \nonumber\\
  P^{p} &=&
  \slashed{q}_1 {\rm{i}} g_p \gamma_5 \left(\slashed{q}_2+m\right)
  \,, \nonumber\\
  P^{\mu\nu}_{t,j} &=&
  \slashed{q}_1\left[
    g^t_{1,j} \frac{i}{2} \sigma^{\mu\nu} +
    g^t_{2,j} \frac{q_1^{\mu}\gamma^\nu - q_1^{\nu}\gamma^\mu}{m} +
    g^t_{3,j} \frac{q_2^{\mu}\gamma^\nu - q_2^{\nu}\gamma^\mu}{m} +
    g^t_{4,j} \frac{q_1^{\mu}q_2^\nu - q_1^{\nu}q_2^\mu}{{m^2}}
    \right] \left(\slashed{q}_2+m\right)
  , \nonumber\\
\end{eqnarray}
with $p=q_1+q_2$, $q=q_1-q_2$, $i=1,2,3$ and $j=1,\ldots,4$.
The coefficients are functions of $m$, $s$ and $\epsilon$
and read
\begin{align}
  & g^v_{1,1} = \frac{s}{4(1-\ep)(s-m^2)^2}\,,&&
  & g^v_{2,1} = -\frac{(-3 + 2\ep)m^2s}{4(1-\ep)(s-m^2)^3} \,,\nonumber\\
  & g^v_{3,1} = -\frac{m^2(-2m^2 + 2\ep m^2- s)}{{4}(1-\ep)(s-m^2)^3}\,,&&
\end{align}
\begin{align}
  & g^v_{1,2} = {-} \frac{m^2}{4(1-\ep)(s-m^2)^2}\,,&&
  & g^v_{2,2} = \frac{m^2(-m^2 - 2s + 2\ep s)}{4(1-\ep)(s-m^2)^3} \,,\nonumber\\
  & g^v_{3,2} = \frac{(-3 + 2\ep)m^4}{{4}(1-\ep)(s-m^2)^3}\,,&&
\end{align}
\begin{align}
  & g^v_{1,3} = \frac{m^2}{8(1-\ep)(s-m^2)^2}\,,&&
  & g^v_{2,3} = -\frac{(-3 + 2\ep)m^4}{8(1-\ep)(s-m^2)^3} \,,\nonumber\\
  & g^v_{3,3} = -\frac{m^2(-5m^2 + 4\ep m^2 + 2s - 2\ep s)}{{8}(1-\ep)(s-m^2)^3}\,,&&
\end{align}
\begin{align}
  & g^a_{1,1} = \frac{s}{4(1-\ep)(s-m^2)^2}\,,&&
  & g^a_{2,1} = \frac{(-3 + 2\ep)m^2 s}{{4}(1-\ep)(s-m^2)^3}\,, \nonumber\\
  & g^a_{3,1} = \frac{m^2(-2m^2 + 2\ep m^2 - s)}{4(1-\ep)(s-m^2)^3}
  \,,&&
\end{align}
\begin{align}
  & g^a_{1,2} = -\frac{m^2}{4(1-\ep)(s-m^2)^2}\,,&&
  & g^a_{2,2} = -\frac{m^2(-m^2 - 2s + 2\ep s)}{{4}(1-\ep)(s-m^2)^3}
  \,,\nonumber\\
  & g^a_{3,2} = -\frac{(-3 + 2\ep)m^4}{4(1-\ep)(s-m^2)^3}
  \,,&&
\end{align}
\begin{align}
  & g^a_{1,3} = \frac{m^2}{{8}(1-\ep)(s-m^2)^2}\,,&&
  & g^a_{2,3} = \frac{(-3 + 2\ep)m^4}{8(1-\ep)(s-m^2)^3}
  \,,\nonumber\\
  & g^a_{3,3} = \frac{m^2(-5m^2 + 4\ep m^2 + 2s - 2\ep s)}{{8}(1-\ep)(s-m^2)^3}
  \,,&&
\end{align}
\begin{align}
  & g^s = \frac{1}{2(m^2- s)}\,,&&
\end{align}
\begin{align}
  & g^p = \frac{1}{2({m^2-s})}\,,&&
\end{align}

\begin{align}
  &g^t_{1,1} = -\frac{1}{2(1 - 3\ep + 2\ep^2)(m^2 - s)}\,,&&
  g^t_{2,1} = \frac{m^2}{2(-1+\ep)(-1+2\ep)(m^2 - s)^2}\,,\nonumber\\
  &g^t_{3,1} = 0\,,&&
  g^t_{4,1} = -\frac{m^2}{2(1 - 3\ep + 2\ep^2)(m^2 - s)^2}\,,
\end{align}
\begin{align}
  &g^t_{1,2} = -\frac{m^2}{ ((-1+\ep)(-1+2\ep)(m^2 - s)^2}\,,&&
  g^t_{2,2} = \frac{ -(-3 + 2\ep)m^4 }{2(-1 + \ep)(-1+2\ep)(m^2 - s)^3}\,,\nonumber\\
  &g^t_{3,2} = \frac{m^2}{4(-1 + \ep)(m^2 - s)^2}\,,&&
  g^t_{4,2} = -\frac{(-3 + 2\ep)m^4}{2(-1 + \ep)(-1 + 2\ep)(m^2 - s)^3}\,,
\end{align}
\begin{align}
  &g^t_{1,3} = 0\,,&&
  g^t_{2,3} = \frac{m^2}{4(-1 + \ep)(m^2 - s)^2 }\,,\nonumber\\
  &g^t_{3,3} = 0\,,&&
  g^t_{4,3} = 0\,,
\end{align}
\begin{align}
  &g^t_{1,4} = \frac{m^2}{(1 - 3\ep + 2\ep^2)(m^2 - s)^2}\,,&&
  g^t_{2,4} = \frac{(-3 + 2\ep)m^4}{ 2(1-3\ep+2\ep^2)(m^2 - s)^3 }\,,\nonumber\\
  &g^t_{3,4} = 0\,,&&
  g^t_{4,4} = -\frac{\ep(-3+2\ep)m^4}{(1 - 3\ep + 2\ep^2)(m^2 - s)^3}\,.
\end{align}


\section{\label{app::code}Implementation in computer code}

In this appendix we present the implementation of the three-loop
form factors for the heavy-to-light transition in the Fortran library \texttt{FFh2l}.
The library numerically evaluates the third-order corrections to the form factors.
The code is deposited on Zenodo~\cite{fael_2024_11046426} and also available at the web address
\begin{verbatim}
https://gitlab.com/formfactors3l/ffh2l
\end{verbatim}
where documentation and sample programs can be found. The code provides interpolation grids
and series expansion which can be used for instance in a Monte Carlo program.

We do not implement all series expansion presented in Eq.~\eqref{eq::som2},
instead we use Chebyshev interpolation grids in the range
$-75< s/m^2 < 15/16$.
Around the singular points $s/m^2 = 1, - \infty $ we implement the power-log expansions.

The Fortran library \texttt{FFh2l} can be cloned from Gitlab with
\begin{verbatim}
$ git clone https://gitlab.com/formfactors3l/ffh2l.git
\end{verbatim}
A Fortran compiler such as \verb|gfortran| is required.
The library can be compiled by running
\begin{verbatim}
$ ./configure
make
\end{verbatim}
Running \verb|make| without further arguments generates the static library \verb|libffh2l.a|
which can be linked to the user’s program.  The module files
are located in the directory \texttt{modules}. They must be also passed to the compiler.
This gives access to the public functions and
subroutines. The names of all subroutines start with the suffix \verb|ffh2l_|.

In order to explain the functionality of the library, let us analyze the
following sample program which evaluates the vector form factor at three-loops.
\begin{verbatim}
program example1
  use ffh2l
  implicit none

  double complex :: ff
  double precision :: s = 0.3d0
  integer :: eporder

  print *,"EXAMPLE 1: Numerical evaluation of"
  print *,"the vector form factor F1 at s = 3/10"
  print *,"---------------------------------"
  print *,"Default configuration:"
  print *,"  - nl =4"
  print *,"  - nh =1"
  print *,""

  do eporder = -6,0
    print *,"F1( s = ",s,", ep = ",eporder," ) = ", ffh2l_veF1(s,eporder)
  enddo

  print *,""
  print *,"Form factor: finite remainder after IR subtraction"
  print *,"F1^fin( s = ",s,") = ", ffh2l_veF1_fin(s)

end program example1
\end{verbatim}
In the preamble of the program, one includes \verb|use ffh2l| to load the respective module.
The form factor is computed by the function \verb|ffh2l_veF1(s,eporder)| which returns
the corresponding order in $\epsilon$ of the ultraviolet-renormalized (but not infrared subtracted) form factor
$F_1^{v,(3)}$.
The result is the third-order correction in the expansion parameter $\alpha_s^{(n_l)}/(4\pi)$,
the strong coupling constant renormalized in the $\overline{\mathrm{MS}}$ scheme
with the renormalization scale set to the heavy-quark mass: $\mu=m$.

For the other form factors, the user can replace \verb|veF1| in the function name with one of the following:
\verb|veF2, veF3, axF1, axF2, axF3, scF1, psF1, teF1, teF2, teF3, teF4|.
Note that the form factors \verb|scF1| and \verb|psF1|
have been implemented using as renormalization constants for the
currents $Z_s =Z_p = Z^{\overline{\mathrm{MS}}}_m$.
In addition to the 12 routines aforementioned,
the user can utilize \verb|scF1OS| and \verb|psF1OS| to obtain results for the scalar and pseudoscalar form factors
with $Z_s =Z_p = Z^{\mathrm{OS}}_m$ for the
current renormalization.

The functions return a \verb|double complex| and have the following two inputs:
\begin{verbatim}
double complex function ffhl2_veF1 (s,eporder)
    double precision, intent(in) :: s
    integer, intent(in) :: eporder
\end{verbatim}
The variable \verb|s| is the value of the momentum transfer normalized w.r.t.\ the squared quark mass.
The order in the dimensional regulator $\epsilon=(4-d)/2$ is set by the integer \verb|eporder|.
Only the values \verb|eporder=-6,...,0| are valid.
These form factors still contains poles since we do not perform the infrared subtraction.
In this way, any infrared subtraction scheme can be applied and it is the task of the user to implement it.

For completeness, we also implement the finite remainder at three loops after minimal subtraction of the infrared poles,
as described in section~\ref{sec::IR}.
In the example above, the finite remainder for the
vector form factor $F_{1}^{v,(3)}$ is obtained using the function \verb|ffh2l_veF1_fin(s)|.
It returns the third-order corrections in the expansion parameter $\alpha_s^{(n_l)}/(4\pi)$.
Here the strong coupling constant is renormalized in the $\overline{\mathrm{MS}}$ scheme
with the renormalization scale $\mu=m$.
The finite remainders for the other form factors are obtained substituting \verb|veF1|
with one of the following:
\verb|veF2, veF3, axF1, axF2, axF3, scF1, psF1, teF1, teF2, teF3, teF4|.
Also in this case, the routines with \verb|scF1| and \verb|psF1|
correspond to the form factors renormalized with
$Z_s =Z_p = Z^{\overline{\mathrm{MS}}}_m$.
We provide additionally two routines identified by
\verb|scF1OS| and \verb|psF1OS| for the finite remainder
of the scalar and pseudoscalar form factors with $Z_s =Z_p = Z^{\mathrm{OS}}_m$.

Each function returns a \verb|double complex| and has the following two inputs:
\begin{verbatim}
double complex function ffhl2_veF1_fin (s)
    double precision, intent(in) :: s
\end{verbatim}
The variable \verb|s| is the value of the momentum transfer normalized w.r.t.\ the squared heavy-quark mass.

In the current implementation, the numerical values of the Casimir are hard coded for QCD
in the file \verb|ffh2l_global.F90|. We set $C_F=4/3, C_A=3, T_F=1/2$.
By default the number of massless and massive quarks are set to $n_l = 4$ and $n_h = 1$, respectively.
The user can modify the values, for instance $n_l=3$ and $n_h=0$, in the following
way:
\begin{verbatim}
integer :: nl = 3
integer :: nh = 0

call ffh2l_set_nl(nl)
call ffh2l_set_nh(nh)
\end{verbatim}

In addition to the Fortran library, we provide also a Mathematica interface
by making use of Wolfram’s MathLink interface (for details on the setup see Ref.~\cite{Hahn:2011gf}).
The interface provides a convenient tool for numerical evaluation and cross-check of our results within Mathematica.
The interface is complied with
\begin{verbatim}
$ make mathlink
\end{verbatim}
To use the library within Mathematica, the interface must be loaded:
\begin{verbatim}
In[] := Install["PATH/ffh2l"]
\end{verbatim}
where \verb|PATH/ffh2l| is the location where the MathLink executable \verb|ffh2l| is saved.
The ultraviolet renormalized form factors in QCD are evaluated with a call to one of the following functions:
\begin{verbatim}
FFh2lveF1   FFh2lveF2   FFh2lveF3
FFh2laxF1   FFh2laxF2   FFh2laxF3
FFh2lscF1   FFh2lscF1OS
FFh2lpsF1.  FFh2lpsF1OS
FFh2lteF1   FFh2lteF2   FFh2lteF3   FFh2lteF4
\end{verbatim}
For instance, the order $\epsilon^0$ in the ultraviolet-renormalized form factor $F_1^{v,(3)}$
is obtained with the following command
\begin{verbatim}
In[] := s = 3/10;
In[] := eporder = 0;
In[] := FFh2lveF1[s,eporder]
Out[]:= 2439.87
\end{verbatim}
The finite remainders of the form factors after infrared subtraction
are obtained by calling the functions
\begin{verbatim}
FFh2lveF1Fin   FFh2lveF2Fin   FFh2lveF3Fin
FFh2laxF1Fin   FFh2laxF2Fin   FFh2laxF3Fin
FFh2lscF1Fin   FFh2lscF1OSFin
FFh2lpsF1Fin   FFh2lpsF1OSFin
FFh2lteF1Fin   FFh2lteF2Fin   FFh2lteF3Fin   FFh2lteF4Fin
\end{verbatim}
For example, the finite remainder of $F_1^{v,(3)}$ is calculated with
\begin{verbatim}
In[] := s = 3/10;
In[] := FFh2lveF1Fin[s]
Out[]:= -8467.54
\end{verbatim}
Also in Mathematica, it is possible to modify the default values of $n_l$ and $n_h$ in the following way:
\begin{verbatim}
In[] := nl=3;
In[] := nh=0;
In[] := FFh2lSetNl[nl]
In[] := FFh2lSetNh[nh]
\end{verbatim}

\end{appendix}

\bibliographystyle{jhep}
\bibliography{main.bib}

\end{document}